\documentclass[journal]{IEEEtran}
\usepackage[cmex10]{amsmath}
\usepackage{amsfonts}
\usepackage{amssymb}
\usepackage{algorithmic}
\usepackage{algorithm}
\usepackage{array}
\usepackage{bm}

\usepackage{textcomp}
\usepackage{stfloats}
\usepackage{url}
\usepackage{verbatim}
\usepackage{graphicx}
\usepackage{cite}

\usepackage[subtle,tracking=normal]{savetrees}
\usepackage{soul}
\usepackage[dvipsnames]{xcolor}
\sethlcolor{CornflowerBlue}

\usepackage{multirow}

\usepackage[tight,footnotesize]{subfigure}
\usepackage{subfiles}
\usepackage{pdfpages}
\usepackage{tikz}
\usetikzlibrary {arrows.meta}

\newtheorem{proposition}{Proposition}
\newtheorem{remark}{Remark}

\begin{document}

\title{Optimal and Suboptimal Decoders under\\ Finite-Alphabet Interference:\\ A Mismatched Decoding Perspective}

\author{
    \IEEEauthorblockN{
        Sibo~Zhang,~\IEEEmembership{Graduate Student Member,~IEEE}, Bruno~Clerckx,~\IEEEmembership{Fellow,~IEEE}}
    \thanks{S. Zhang is with the Department of Electrical and Electronic Engineering at Imperial College London, London SW7 2AZ, UK and BBC Research and Development, The Lighthouse, White City Place, 201 Wood Lane, London, W12 7TQ, U.K. (e-mail: sibo.zhang19@imperial.ac.uk).}
    \thanks{B. Clerckx is with the Department of Electrical and Electronic Engineering at Imperial College London, London SW7 2AZ, UK (e-mail: b.clerckx@imperial.ac.uk).}
    \thanks{This work was supported in part by the UK Engineering and Physical Sciences Research Council, Industrial Case award number 210163.}

    }



\maketitle

\begin{abstract}
Interference widely exists in communication systems and is often not optimally treated at the receivers due to limited knowledge and/or computational burden. Evolutions of receivers have been proposed to balance complexity and spectral efficiency, for example, for 6G, while commonly used performance metrics, such as capacity and mutual information (MI), fail to capture the suboptimal treatment of interference, leading to potentially inaccurate performance evaluations. Mismatched decoding is an information-theoretic tool for analyzing communications with suboptimal decoders. In this work, we use mismatched decoding to analyze communications with decoders that treat interference suboptimally, aiming at more accurate performance metrics. Specifically, we consider a finite-alphabet input Gaussian channel under interference, representative of modern systems, where the decoder can be matched (optimal) or mismatched (suboptimal) to the channel. The matched capacity is derived using MI, while a lower bound on the mismatched capacity under various decoding metrics is derived using generalized mutual information (GMI). We show that the decoding metric in the proposed channel model is closely related to the behavior of the demodulator in bit-interleaved coded modulation (BICM) systems. Simulations illustrate that GMI/MI accurately predicts the throughput of BICM-type systems {with various demodulators}. Finally, we extend the channel model and the GMI to multiple antenna cases, with an example of multi-user multiple-input-single-output (MU-MISO) precoder optimization problem considering GMI under different decoding strategies. In short, this work discovers new insights about the impact of interference, proposes novel receivers, and introduces a new design and performance evaluation framework that more accurately captures the effect of interference.
\end{abstract}

\section{Introduction}

Shannon's channel coding theorem provides limits on reliable communication rates subject to optimal encoding and decoding \cite{Shannon,Gallager}. The channel coding theorem is often proved by assuming {decoding techniques that rely on and optimally exploit the channel transition law, such as joint typicality decoding or maximum-likelihood decoding}. However, in practice, it is often the case that suboptimal decoding is implemented because, for example, the channel knowledge is not sufficient at the decoder, or there are constraints on the decoding technique in use, possibly regarding computational burden. 

Mismatched decoding \cite{Mismatched_1,Lapidoth_1,Albert} is a theory that addresses the problem of reliable communications subject to suboptimal decoders. Mismatched capacity, i.e., the supremum of rates that allow reliable communications under mismatched decoding, is often a difficult problem, because this would require the codebook to be optimized according not only to the channel knowledge but also to the decoding rule. The common approach is to make certain assumptions on the coding ensemble (i.e., {the random process that generates the codebook}), often suboptimal, and study the resulting achievable rates. This approach often leads to lower bounds of mismatched capacity. Two well-known lower bounds of mismatched capacity are generalized mutual information (GMI) and LM rate,\footnote{The LM rate appears to have originated in \cite{Mismatched_1}, implicitly denoting ``Lower bound on the Mismatch capacity''.} where the GMI assumes i.i.d. random coding and the LM rate assumes constant-composition random coding \cite{Albert}. 

Mismatched decoding has been evaluated under various scenarios. One motivation behind using mismatched decoders is often from the concern on decoder's complexity. For example, \cite{Lapidoth_2} addresses nearest neighbour decoding under non-Gaussian channels, as maximum-likelihood decoding for non-Gaussian channels can be computationally complicated depending on the noise distribution. \cite{Shamai_2} investigates decoders under constraints of finite precision arithmetic. \cite{Lapidoth_3} considers multiple access channels with simplified decoders. In \cite{Fabregas_2}, bit-interleaved coded modulation (BICM) systems are analyzed using mismatched decoding. \cite{Jonathan_MAC_TIT} studies multiple access channels with mismatches in successive decoding.

The above examples consider mismatched decoding due to decoder limitations. For wireless systems, fluctuating channels and limited resource for channel estimation bring about the other common cause of mismatched decoding, namely imperfect channel knowledge. \cite{Lapidoth_4} addresses achievable rate with imperfect side information at the receiver. \cite{Shamai_1} and \cite{Fabregas_1} consider Multiple-Input-Multiple-Output (MIMO) systems under block fading channels and imperfect Channel State Information at Receiver. \cite{Goldsmith} studies the scenario where imperfect channel knowledge exists at both the transmitter and receiver.

Interference is a common challenge in wireless communication systems and is addressed in various aspects of the system design. For cellular networks, in 3GPP Release 18, multiple work items address interference from different perspectives \cite{3gpp_rel18_WI}. Among others, advanced MIMO receiver designs that suppress intra- and inter-cell interference are studied \cite{3gpp_NR_demod}. 
{\cite{Industrial6G} provides viewpoints on potential 6G RAN technologies, where the impact of interference is highlighted in various aspects.} 
To understand the impact of interference, the vast majority of work in the literature models both input signal and interference using Gaussian distributions, although most systems apply finite-alphabet signals. Such approaches lead to acceptable approximations of system performance under the impact of interference and benefit from the simple analytical achievable rate expressions. More precise but computationally more complicated evaluation of achievable rates can be obtained using constellation-constrained mutual information (MI) \cite{Wu_survey}. However, both approaches assume optimal decoders, which can be difficult in practice due to limited interference knowledge and the complexity issue. The impact of suboptimal decoders when dealing with finite-alphabet interference remains to be explored.

In this work, we investigate the effect of finite-alphabet interference under optimal and suboptimal decoders. The contribution is as follows:
\begin{enumerate}
    \item We define a special channel model, namely Finite-Alphabet Gaussian Channel under Interference (FAGCI), which is a single-input-single-output (SISO) channel with finite-alphabet input, finite-alphabet interference and Gaussian noise. This channel model is to reflect practical scenarios where the communication channel suffers from not only noise but also interference from co-scheduled users or other communication links. Maximum-likelihood decoding under FAGCI is given, but can be computationally complicated once the interference alphabet has a high cardinality. A more practical decoding approach can be to treat part of the interference as Gaussian random variables (RV). We derive the MI and GMI of FAGCI, which correspond to the achievable rates under optimal and suboptimal decoders. Through analytical properties of the GMI/MI and numerical simulations, insights on the impact of finite-alphabet interference under different decoders are obtained.

    \item Noting the difference in computational complexity between the matched decoding metric and mismatched decoding metrics, we propose two novel decoding metrics, namely generalized Gaussian decoding metric and interference decomposition decoding metric, whose computational complexity remains low. The numerical results show that they lead to improved achievable rates compared to decoders that treat interference as a Gaussian RV. 

    \item We show that the notion of decoding metric in the FAGCI model is related to BICM-demodulators, and propose low-complexity demodulators based on the proposed decoding metrics. Through link-level simulations (LLS), it is shown that the proposed demodulators lead to improved throughput. We also show a strong correlation between the GMI and throughput when the decoding metric and demodulator apply the same principle for treating interference. Since computing the GMI is much more time-efficient than conducting LLS, such a correlation indicates that GMI can be an important performance indicator for BICM systems.

    \item Motivated by the precise performance evaluation given by the GMI and the correlation between the GMI and throughput, we further consider signal processing based on the GMI. In particular, we extend the GMI results for FAGCI to multiple-antenna cases, and provide an example of multi-user multiple-input-single-output (MU-MISO) system, where each user potentially suffers from inter-user interference. We propose optimizing the precoders by taking the GMI as the objective function. Therefore, such an optimization problem incorporates decoding strategies on the user side. An optimization algorithm is proposed, and the numerical results reflect the impact of suboptimal decoders on the achievable rate and the necessity of using GMI as the objective function. The imperfection in receivers on treating finite-alphabet interference widely exists in practice, yet means of incorporating such imperfection is mostly overlooked in signal processing. The proposed method pioneers a novel system design framework that addresses this problem.
    
\end{enumerate}

The following is a brief summary of the new insights and implications of finite-alphabet interference derived from this work.
\begin{enumerate}
    \item The achievable rate under finite-alphabet input signal and interference is in general non-monotonic w.r.t. input power.
    \item Different approaches to addressing interference during decoding result in significant variations in the achievable rate.
    \item Strong interference is not always detrimental: under proper interference treatment, the achievable rate under very strong interference is as if that interference were not present.
    \item Decoding metrics and BICM-demodulators are strongly related, and innovations in the former can inspire improvements in the latter.
    \item {The GMI/MI reflects the physical-layer throughput under different BICM demodulators. Therefore, it can serve as both a computationally efficient tool for performance evaluation of demodulators and an accurate physical-layer abstraction for signal processing and system-level simulations (SLS).}
    \item Discrepancy between the decoding metric and the objective function in signal processing can be harmful to system performance.
\end{enumerate}

\emph{Organization:} The rest of the paper is organized as follows. In Section II, we propose the FAGCI model, introduce assumptions on the encoder and decoder, the resulting GMI and its approximation, and present numerical results from the GMI along with some new insights. In Section III, two improved decoding metrics for the FAGCI model are proposed, namely the generalized Gaussian decoding metric and the interference decomposition decoding metric, followed by numerical results on the GMI led by them. In Section IV, we show the connection between the decoding metric in the FAGCI model and the BICM-demodulators, propose low-complexity demodulators based on the proposed decoding metrics, and provide numerical results comparing the GMI and the throughput performance of the demodulators. Section V provides an example of precoder optimization in MU-MISO systems utilizing the GMI, which pioneers a novel system design framework that captures receiver imperfection in treating finite-alphabet interference. Finally, Section VI concludes this paper.

\emph{Notations:} Scalars are denoted by normal letters.
Sequences and vectors are denoted by bold letters.
Random variables are denoted by capital letters, while the specific values are denoted by lower case letters. {Bold uppercase letters can denote matrices or a vector of random variables, depending on the context.} For example, $x$ is an observation of $X$, and $\mathbf{x}$ is an observation of $\mathbf{X}$. Calligraphic letters denote multisets. $P_{\cdot}(\cdot)$, $P_{\cdot\mid \cdot}(\cdot\mid \cdot)$, $f_{\cdot}(\cdot)$ and $f_{\cdot \mid \cdot}(\cdot \mid \cdot)$ denote probability, conditional probability, probability density, and conditional probability density respectively. $\mathbf{I}$ and $\mathbf{0}$ denotes the identity matrix and the zero matrix, whose dimensions are given by their superscripts. $[\cdot]_{m,n}$ denotes the $(m,n)$-th entry of a matrix. $(\cdot)^T$, $(\cdot)^H$, $\| \cdot \|$ and $\| \cdot \|_\text{F}$ denote respectively the transpose, conjugate transpose, Euclidean norm and Frobenius norm of the input entity. $|\cdot|$ denotes the absolute value if the argument is a scalar, or the cardinality if the argument is a multiset. $\times$ denotes the Cartesian product of two sets. {$\mathbb{E}_X[\cdot]$ denotes expectation w.r.t. $X$, and the subscript is omitted when unambiguous.} $\mathcal{CN}(\bm{\mu},\mathbf{\Sigma})$ denotes circular symmetric complex Gaussian distribution for vectors whose mean and covariance matrix are $\bm{\mu}$ and $\mathbf{\Sigma}$. $\text{dim}(\cdot)$ denotes the dimension of the argument. {All rate expressions are in units of nats for clarity, while the numerical results are presented in bits.}

\section{Mismatched Decoding for Finite Alphabets}\label{Mismatched Decoding for Finite Alphabets}
\subsection{Finite-Alphabet Gaussian Channel under Interference}\label{section: 2.1}
We consider coded communications with finite-alphabet inputs under additive interference and additive Gaussian noise, or FAGCI. {WLOG}, we assume that there are two interference sources. Thus, at time instance $t$, the received signal can be expressed as
\begin{equation}\label{FAG channel}
    y_t = x_t + i_t + j_t + z_t,
\end{equation}
where $x_t$ and $y_t$ are the complex-valued channel input and output, $i_t$ and $j_t$ are the two complex-valued interference observations, and $z_t$ is the noise observation. $x_t$, $i_t$ and $j_t$ are uniformly drawn from alphabets $\mathcal{X}$, $\mathcal{I}$ and $\mathcal{J}$ correspondingly,\footnote{This work focus on uniformly distributed inputs and interference for clarity. {The extension to non-uniform distributions is straightforward and is briefly addressed at the end of Appendix \ref{Appendix: Proof of mismatched capacity}.}} while $z_t$ are samples of $Z_t\sim\mathcal{CN}(0,\;\sigma_z^2)$. Unless any of $\mathcal{X}$, $\mathcal{I}$ and $\mathcal{J}$ is $\{0\}$, $\mathbb{E}\left[|X_t|^2\right] = \sigma_x^2$, $\mathbb{E}\left[|I_t|^2\right] = \sigma_i^2$ and $\mathbb{E}\left[|J_t|^2\right] = \sigma_j^2$. {$X_t$, $I_t$, $J_t$ and $Z_t$ and are mutually independent are i.i.d. over $t$, leading to a memoryless channel, i.e.,}
\begin{equation}
    f_{\mathbf{Y} \mid \mathbf{X}}(\mathbf{y} \mid \mathbf{x}) = \prod_{t=1}^n f_{Y \mid X}(y_t \mid x_t),
\end{equation}
where $\mathbf{x}$ and $\mathbf{y}$ are the input and output sequences of length $n$. Henceforth, the time index is omitted if no ambiguity arises. 

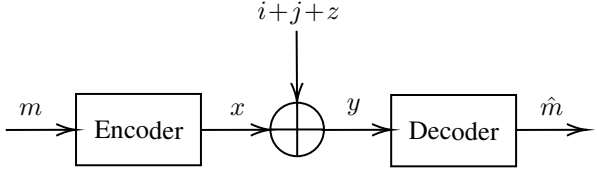
\begin{figure}
    \centering
    \tikzset{every picture/.style={line width=0.75pt}} 

\begin{tikzpicture}[x=0.75pt,y=0.75pt,yscale=-0.9,xscale=0.9]

\draw   (149.4,70.4) -- (219.4,70.4) -- (219.4,110.4) -- (149.4,110.4) -- cycle ;
\draw   (326,71) -- (396,71) -- (396,111) -- (326,111) -- cycle ;
\draw[->]    (396.17,90.5) -- (433.42,90.5) ;

\draw[->]    (110.01,90.5) -- (149.4,90.5) ;

\draw[->]    (219,90.5) -- (258.33,90.5) ;

\draw[->]    (288.33,90.5) -- (326,90.5) ;

\draw   (258.33,90.5) .. controls (258.33,82.05) and (265.05,75.33) .. (273.33,75.33) .. controls (281.62,75.33) and (288.33,82.05) .. (288.33,90.5) .. controls (288.33,98.62) and (281.62,105.33) .. (273.33,105.33) .. controls (265.05,105.33) and (258.33,98.62) .. (258.33,90.5) -- cycle ; \draw   (258.33,90.5) -- (288.33,90.5) ; \draw   (273.33,75.33) -- (273.33,105.33) ;
\draw[->]    (273.33,35) -- (273.33,75.33) ;

\draw (184.4,90.4) node  [rotate=-0.25] [align=left] {Encoder};
\draw (361,91) node  [rotate=-0.25] [align=left] {Decoder};
\draw (274,33) node [anchor=south] [inner sep=0.75pt]   [align=left] {$\displaystyle i+j+z$};
\draw (240,84) node [anchor=south] [inner sep=0.75pt]   [align=left] {$\displaystyle x$};
\draw (305,84) node [anchor=south] [inner sep=0.75pt]   [align=left] {$\displaystyle y$};
\draw (124,84) node [anchor=south] [inner sep=0.75pt]   [align=left] {$\displaystyle m$};
\draw (416,84) node [anchor=south] [inner sep=0.75pt]   [align=left] {$\displaystyle \hat{m}$};

\end{tikzpicture}
\caption{Channel model with interference and noise.}
\end{figure}

\begin{remark}
Note that (\ref{FAG channel}) can represent channels under an arbitrary number of interference sources. Channels under one interference source can be represented by having either $\mathcal{I}$ or $\mathcal{J}$ equal to $\{0\}$. For more than two interference sources, one can have 
\begin{equation}
    i_t = \sum_p i_t^p \text{\;,\;} j_t = \sum_q j_t^q \text{\;with\;} \mathcal{I} = \sum_p \mathcal{I}^p \text{\;and\;} \mathcal{J} = \sum_q \mathcal{J}^q,
\end{equation}
where, for arbitrary range of $p$ and $q$, $i_t^p$ and $j_t^q$ represents the elementary interference observations drawn from $\mathcal{I}^p$ and $\mathcal{J}^q$. We keep two interference sources rather than one because in the following we are interested in scenarios where the two interference sources are treated differently during decoding.
\end{remark}

\subsection{Encoding and Decoding}
We aim at reliably transmitting a message, $m$, over $n$  channel uses, where $m$ is drawn randomly and uniformly from $\{1,\;...\;,\;M\}$. The encoding scheme is given by a codebook $\mathcal{C}=\{\mathbf{x}^{(1)},\;...\;,\;\mathbf{x}^{(M)}\}\subset\mathcal{X}^n$, i.e., to transmit the message $m$, the corresponding codeword $\mathbf{x}^{(m)}$ is transmitted. We further assume i.i.d. random coding ensemble, hence
\begin{equation}\label{random_coding_ensemble}
    P_\mathbf{X}(\mathbf{x}^{(m)}) = \prod_{t=1}^{n} P_X(x_t^{(m)}),
\end{equation}
where $x_t^{(m)}$ is the $t$-th symbol of $\mathbf{x}^{(m)}$, and $n$ is the length of codewords.

The decoder is represented by
\begin{equation}
    \hat{m} = \underset{{l=1,...,M}}{\arg\; \max\;\;\;\;} q^n(\mathbf{x}^{(l)},\;\mathbf{y}), \text{ with } q^n(\mathbf{x}^{(l)},\;\mathbf{y}) = \prod_{t=1}^n q(x^{(l)}_t,\;y_t),
\end{equation}
where $q(x,\;y)$ is a non-negative decoding metric.

For the FAGCI channel model described above, the optimal (matched) decoder maximizes the likelihood, which leads to the following optimal decoding metric:
\begin{equation}\label{matched decoding metric}
    q^\star(x,\;y) = \sum_{\Bar{i}\in\mathcal{I}} \sum_{\Bar{j}\in\mathcal{J}} \exp\left({-\frac{|y-\Bar{i}-\Bar{j}-x|^2}{\sigma_z^2}}\right).
\end{equation}
We are also interested in suboptimal (mismatched) but simpler decoding metrics. Primarily, the following two are considered:
\begin{enumerate}
    \item Partially treat interference as Gaussian RV by treating $I$ optimally and treating $J$ as a Gaussian RV\footnote{{Another interpretation of (\ref{mismatched decoding metric partial}) is to approximate $P_{I+J}(i+j)$ with a Gaussian mixture model.}}:
    \begin{equation}\label{mismatched decoding metric partial}
    q(x,\;y) = \sum_{\Bar{i}\in\mathcal{I}} \exp\left({-\frac{|y-\Bar{i}-x|^2}{\sigma_j^2+\sigma_z^2}}\right).
    \end{equation}
    \item Fully treat interference as a Gaussian RV:
    \begin{equation}\label{mismatched decoding metric full}
    q(x,\;y) = \exp\left({-\frac{|y-x|^2}{\sigma_i^2+\sigma_j^2+\sigma_z^2}}\right).
    \end{equation}
\end{enumerate}
Indeed, if $J \sim \mathcal{CN}(0,\;\sigma_j^2)$ while $I$ remains to be from finite alphabet, (\ref{mismatched decoding metric partial}) is the optimal decoding metric for (\ref{FAG channel}); if $I \sim \mathcal{CN}(0,\;\sigma_i^2)$ and $J \sim \mathcal{CN}(0,\;\sigma_j^2)$, (\ref{mismatched decoding metric full}) is the optimal decoding metric. Note that (\ref{mismatched decoding metric full}) is a special case of (\ref{mismatched decoding metric partial}) as (\ref{mismatched decoding metric partial}) becomes (\ref{mismatched decoding metric full}) when viewing $i$ as $0$, and $j$ as $i+j$.

\subsection{Matched Capacity and a Lower Bound on Mismatched Capacity}
\begin{proposition}[Matched capacity]\label{matched capacity prop}
With optimal decoding, i.e., using (\ref{matched decoding metric}) as the decoding metric, the capacity of (\ref{FAG channel}) is given by the constellation-constrained MI as in (\ref{matched capacity}).
\end{proposition}
\begin{IEEEproof}
    See Appendix \ref{Appendix: Proof of matched capacity}.
\end{IEEEproof}

\begin{proposition}[A lower bound on mismatched capacity]\label{mismatched capacity prop}
With \textbf{any} suboptimal decoding metric denoted by $q(x,\;y)$, the mismatched capacity of (\ref{FAG channel}) is lower bounded by the GMI as in (\ref{mismatched capacity}), which is a function of $q(x,\;y)$. 
\end{proposition}
\begin{IEEEproof}
    See Appendix \ref{Appendix: Proof of mismatched capacity}.
\end{IEEEproof}

\begin{figure*}

\begin{flalign}\label{matched capacity}
    I_{\text{MI}}(\mathcal{X},\;\mathcal{I},\;\mathcal{J},\;\sigma_z^2) = &\log|\mathcal{X}| - \frac{1}{|\mathcal{X}||\mathcal{I}||\mathcal{J}|}\sum_{x\in\mathcal{X},\;i\in\mathcal{I},\; j\in\mathcal{J}}\mathbb{E}\left[\log\sum_{\Bar{x}\in\mathcal{X},\;\Bar{i}\in\mathcal{I},\;\Bar{j}\in\mathcal{J}}
    \exp
    \left({-\frac{|x+i+j+Z-\Bar{x}-\Bar{i}-\Bar{j}|^2}{\sigma_z^2}}\right)\right]&&\nonumber\\
    &+ \frac{1}{|\mathcal{I}||\mathcal{J}|}\sum_{i\in\mathcal{I},\;j\in\mathcal{J}}\mathbb{E}\left[\log\sum_{\Bar{i}\in\mathcal{I},\;\Bar{j}\in\mathcal{J}}\exp\left({-\frac{|i+j+Z-\Bar{i}-\Bar{j}|^2}{\sigma_z^2}}\right) \right]&&
\end{flalign}

\begin{flalign}\label{mismatched capacity}
    I_\text{GMI}(\mathcal{X},\;\mathcal{I},\;\mathcal{J},\;\sigma_z^2)
    =\sup_{s\geq 0}\;\; \log|\mathcal{X}|-\frac{1}{|\mathcal{X}||\mathcal{I}||\mathcal{J}|} \sum_{x\in\mathcal{X},\;i\in\mathcal{I},\;j\in\mathcal{J}}\mathbb{E}\left[ \log \sum_{\Bar{x}\in\mathcal{X}} q(\Bar{x},\;x+i+j+Z)^s - s\log\; q(x,\;x+i+j+Z)\right]&&
\end{flalign}

\begin{flalign}\label{mismatched capacity approx partial}
        I_\text{GMI,partial}^\text{approx.}(\mathcal{X},\;\mathcal{I},\;\mathcal{J},\;\sigma_z^2) 
        =& \log|\mathcal{X}| - \frac{1}{|\mathcal{X}||\mathcal{I}||\mathcal{J}|} \sum_{x\in\mathcal{X},\;i\in\mathcal{I},\;j\in\mathcal{J}} \log \sum_{\Bar{x}\in\mathcal{X},\;\Bar{i}\in\mathcal{I}} \exp\left(-\frac{|x+i+j-\Bar{x}-\Bar{i}|^2}{\sigma_j^2+2\sigma_z^2}\right)&&\nonumber\\
        &+ \frac{1}{|\mathcal{I}||\mathcal{J}|} \sum_{i\in\mathcal{I},\;j\in\mathcal{J}} \log \sum_{\Bar{i}\in\mathcal{I}} \exp\left(-\frac{|i+j-\Bar{i}|^2}{\sigma_j^2+2\sigma_z^2}\right)&&
\end{flalign}

\begin{flalign}\label{mismatched capacity approx full}
        I_\text{GMI,full}^\text{approx.}(\mathcal{X},\;\mathcal{I},\;\mathcal{J},\;\sigma_z^2) 
        =& \log|\mathcal{X}| - \frac{1}{|\mathcal{X}||\mathcal{I}||\mathcal{J}|} \sum_{x\in\mathcal{X},\;i\in\mathcal{I},\;j\in\mathcal{J}} \log \sum_{\Bar{x}\in\mathcal{X}} \exp\left(-\frac{|x+i+j-\Bar{x}|^2}{\sigma_i^2+\sigma_j^2+2\sigma_z^2}\right) + \frac{1}{|\mathcal{I}||\mathcal{J}|} \sum_{i\in\mathcal{I},\;j\in\mathcal{J}} \left(-\frac{|i+j|^2}{\sigma_i^2+\sigma_j^2+2\sigma_z^2}\right)&&
\end{flalign}
\hrule
\end{figure*}

\begin{remark}
    In (\ref{mismatched capacity}), the optimization over $s$ is convex \cite{Albert}.
\end{remark}

\begin{remark}
    The GMI given by (\ref{mismatched capacity}) is the highest achievable rate of (\ref{FAG channel}) under the assumption of i.i.d. random coding ensemble given by (\ref{random_coding_ensemble}). Breaking such assumption may lead to higher achievable rates, for example, by using constant-composition random coding ensemble\cite{Albert}.
\end{remark}

\begin{remark}
     In (\ref{mismatched capacity}), $I_\text{GMI}(\mathcal{X},\;\mathcal{I}+\mathcal{J},\;\{0\},\;Z)$ with $s=1$ and $q(x,\;y)$ from (\ref{matched decoding metric})  equals $I_{\text{MI}}(\mathcal{X},\;\mathcal{I},\;\mathcal{J},\;Z)$, as optimal decoding is a special case of mismatched decoding. 
\end{remark}

\subsection{Approximation of mismatched capacity}
One potential challenge of utilizing (\ref{mismatched capacity}) stems from the fact that it is difficult to compute the expectation term analytically. One may estimate (\ref{mismatched capacity}) empirically at the cost of a high computational burden. For the sake of computation time, we provide a closed-form approximation of (\ref{mismatched capacity}). 

\begin{proposition}[Approximation of {the GMI}]
    With $q(x,y)$ given by (\ref{mismatched decoding metric partial}) or (\ref{mismatched decoding metric full}), (\ref{mismatched capacity}) can be approximated by (\ref{mismatched capacity approx partial}) or (\ref{mismatched capacity approx full}) respectively.
\end{proposition}

\begin{IEEEproof}
    By setting $s=1$ and applying Jensen's inequality to the expectation in (\ref{mismatched capacity}), the expectation can be moved inside the two $\log$ operators. With $q(x,\;y)$ given by (\ref{mismatched decoding metric partial}) or (\ref{mismatched decoding metric full}), the resulting expression is further simplified using Gaussian integral into (\ref{mismatched capacity approx partial}) or (\ref{mismatched capacity approx full}) respectively.
\end{IEEEproof}

\subsection{Numerical results on GMI for FAGCI model}\label{section: 2 results}
We simulate the behavior of GMI and MI in the FAGCI model as expressed by (\ref{FAG channel}) under different scenarios. In the following results, the two aforementioned mismatched decoding metrics are considered: 
\begin{itemize}
    \item Treating only $J$ as a Gaussian RV, as given by (\ref{mismatched decoding metric partial}): this is denoted by ``GMI (partial)'' in the following results.
    \item Treating both $I$ and $J$ as Gaussian RVs, as given by (\ref{mismatched decoding metric full}): this is denoted by ``GMI (full)'' in the following results.
\end{itemize}
We also plot the approximations of GMI, i.e., (\ref{mismatched capacity approx partial}) and (\ref{mismatched capacity approx full}), which are denoted as ``GMI (partial) approx.'' and ``GMI (full) approx.'' respectively. In the following, we assume that $x$ is generated using 16-QAM, while $i$ and $j$ are generated using QPSK.\footnote{{The observations generally hold under different constellation setups.}}

\begin{figure}
    \centering
    \includegraphics[width=6.5cm]{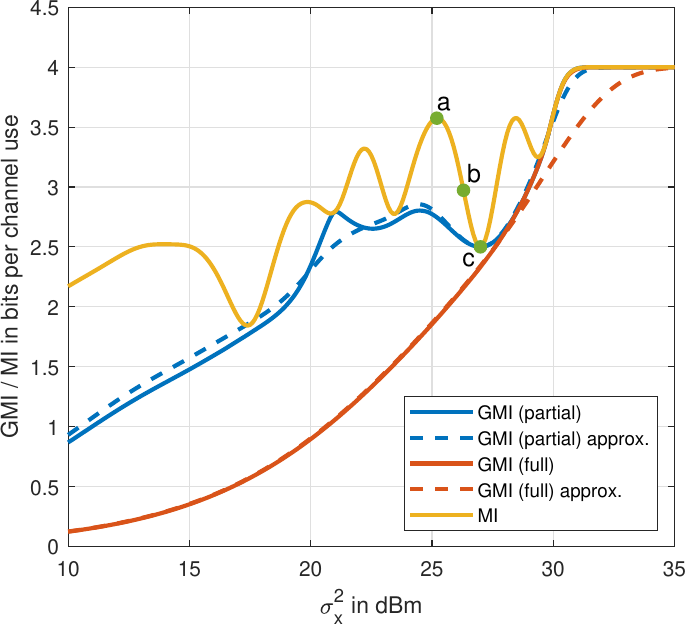}
    \caption{GMI and MI with $\sigma_i^2=20$dBm, $\sigma_j^2=10$dBm and $\sigma_z^2=0$dBm.}
    \label{Fig: GMI_vs_MI_varying_X_pow}
\end{figure}

\begin{figure*}
    \subfigure[$\sigma_x^2 = 25.2$dBm, $H(X,I,J | Y)=0.4324$bits, $I_{\text{MI}}=3.5742$bits.]{\includegraphics[width=5.5cm]{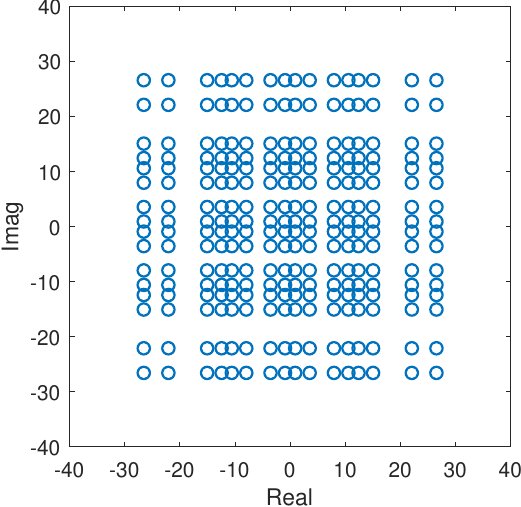}}
    \hspace{0.5cm}
    \subfigure[$\sigma_x^2 = 26.3$dBm, $H(X,I,J | Y)=1.0344$bits, $I_{\text{MI}}=2.9722$bits.]{\includegraphics[width=5.5cm]{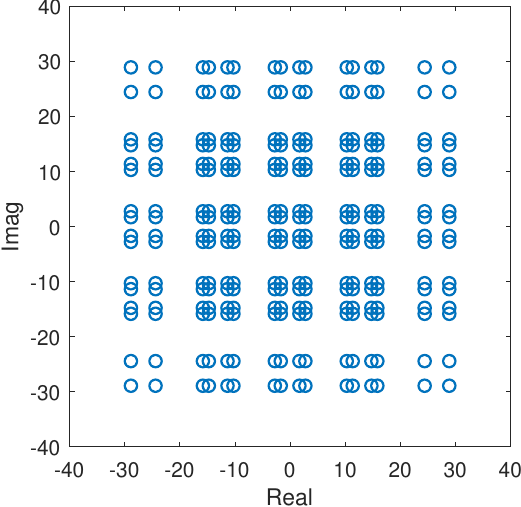}}
    \hspace{0.5cm}
    \subfigure[$\sigma_x^2 = 27$dBm, $H(X,I,J | Y)=1.5064$bits, $I_{\text{MI}}=2.5002$bits.]{\includegraphics[width=5.5cm]{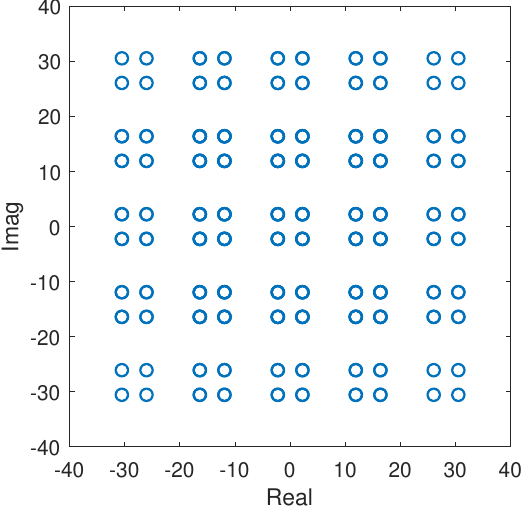}}
    \caption{Visualization of $\mathcal{X+I+J}$ at the three marked points in Figure \ref{Fig: GMI_vs_MI_varying_X_pow}.}
    \label{Fig: sum constellation}
\end{figure*}

In Figure \ref{Fig: GMI_vs_MI_varying_X_pow}, we consider a scenario with fixed interference and noise powers and characterize GMI and MI under different input power, i.e., $\sigma_x^2$. It can be observed that all the considered decoding metrics generally lead to improved achievable rates as $\sigma_x^2$ increases. 
{With very high $\sigma_x^2$, different decoding metrics under consideration lead to the same performance, saturating at 4 bits, which is limited by the cardinality of 16-QAM. }
With limited $\sigma_x^2$, it can be observed that ``MI'' upper-bounds ``GMI (partial)'', and ``GMI (partial)'' upper-bounds ``GMI (full)''. This aligns with expectations, because matched decoding uses optimal decoding metric, which utilizes full knowledge on interference. With mismatched decoding, only part of the interference knowledge is used. (\ref{mismatched decoding metric partial}) uses full knowledge on $I$ but only the variance of $J$ by approximating it as Gaussian RV. (\ref{mismatched decoding metric full}) uses only the variance of $I$ and $J$ by approximating both $I$ and $J$ as Gaussian RVs, hence uses even less knowledge on interference. In most cases, the more interference knowledge is utilized, the more accurate the decoding is. Hence, the order is almost always ``MI'' $\geq$ ``GMI (partial)'' $\geq$ ``GMI (full)''. {When $\sigma_x^2$ is between 28dBm and 29.5dBm, the two GMI metrics become similar, and both are lower than MI. This is because the impact of suboptimally treating $J$ becomes negligible, whereas the effect of suboptimally treating 
$I$ remains significant, as $I$ is stronger than $J$ in the setup. When $\sigma_x^2$ is higher than 29.5dBm, the three rates become similar, because the impact of suboptimally treating both $I$ and $J$ becomes negligible.} Interestingly, MI does not increase monotonically as $\sigma_x^2$ grows but fluctuates. This is because, with finite constellations, the achievable rate also depends on the sum constellations, $\mathcal{X+I+J}$ and $\mathcal{I+J}$. Note that in (\ref{matched capacity}), the second and third terms are conditional entropy, $H(X,\;I,\;J \mid Y)$ and $H(I,\;J \mid Y,\;X)$, respectively. Such conditional entropy depends on the distinguishability of different points in the sum constellations. Since $\sigma_i^2$ and $\sigma_j^2$ are fixed, $H(I,\;J \mid Y,\;X)$ is unchanged (computed to be 0.0066 bits). The MI only depends on $H(X,\;I,\;J \mid Y)$, and therefore the sum constellation, $\mathcal{X+I+J}$. Figure \ref{Fig: sum constellation} visualizes $\mathcal{X+I+J}$ at the three marked points in Figure \ref{Fig: GMI_vs_MI_varying_X_pow}, with the corresponding $H(X,\;I,\;J \mid Y)$ and $I(X \mid Y)$ given in the captions. It can be observed that, in the three examples, as $\sigma_x^2$ increases, the sum constellation becomes less distinguishable, leading to increase in $H(X,\;I,\;J \mid Y)$, and therefore drop in $I(X \mid Y)$.

\begin{figure}
    \centering
    \includegraphics[width=6.5cm]{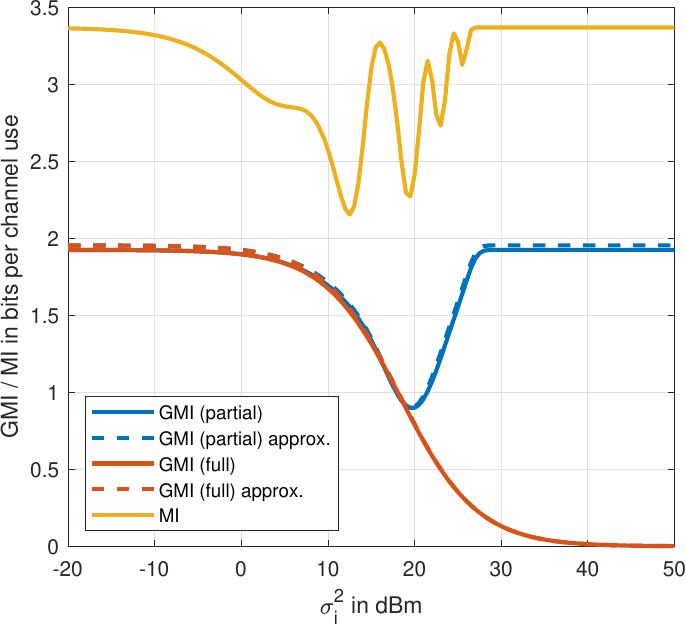}
    \caption{GMI and MI with $\sigma_x^2=20$dBm, $\sigma_j^2=15$dBm and $\sigma_z^2=0$dBm.}
    \label{Fig: GMI_vs_MI_varying_I_pow}
\end{figure}

By fixing $\sigma_x^2$ and $\sigma_j^2$, and adjusting $\sigma_i^2$, Figure \ref{Fig: GMI_vs_MI_varying_I_pow} can be obtained. It can be observed that there is always a gap between the GMI and MI, indicating that the two mismatched decoders always underperform the optimal decoding metric. This gap exists because $J$ with non-negligible power is approximated as a Gaussian RV by the mismatched decoders, while optimally treated by the matched decoder. It can also be observed that, as $\sigma_i^2$ grows, ``GMI (partial)'' first drops, then grows and saturates in the end, while ``GMI (full)'' only decreases until zero. The gap between the two GMI curves is due to the difference in treating $I$, and such a gap increases as $\sigma_i^2$ increases. As for the MI, the fluctuation at the medium $\sigma_i^2$ regime is due to the non-monotonic change in the distinguishability of the sum constellations as analyzed above. Notably, {both the MI and ``GMI (partial)'' saturate to their respective values, which remain the same as} $\sigma_i^2 \rightarrow 0 \text{ or } +\infty$. This trend is formally stated by Proposition \ref{prop: MI_GMI_saturation}. 

\begin{proposition}\label{prop: MI_GMI_saturation}
\begin{equation}
\begin{split}
    \lim_{\sigma_i^2 \rightarrow 0} I_{\text{MI}}(\mathcal{X},\;\mathcal{I},\;\mathcal{J},\;\sigma_z^2)
    =&\lim_{\sigma_i^2 \rightarrow +\infty}  I_{\text{MI}}(\mathcal{X},\;\mathcal{I},\;\mathcal{J},\;\sigma_z^2)\\
    =&I_{\text{MI}}(\mathcal{X},\;\{0\},\;\mathcal{J},\;\sigma_z^2),
\end{split}
\end{equation}
and
\begin{equation}
\begin{split}
    \lim_{\sigma_i^2 \rightarrow 0} I_{\text{GMI}}(\mathcal{X},\;\mathcal{I},\;\mathcal{J},\;\sigma_z^2)
    =&\lim_{\sigma_i^2 \rightarrow +\infty}  I_{\text{GMI}}(\mathcal{X},\;\mathcal{I},\;\mathcal{J},\;\sigma_z^2)\\
    =&I_{\text{GMI}}(\mathcal{X},\;\{0\},\;\mathcal{J},\;\sigma_z^2)
\end{split}
\end{equation}
with $q(x,\;y)$ given by (\ref{mismatched decoding metric partial}).

\end{proposition}
\begin{IEEEproof}
    See Appendix \ref{Appendix: Proof of MI/GMI saturation}.
\end{IEEEproof}

\begin{figure}
    \centering
    \includegraphics[width=6.5cm]{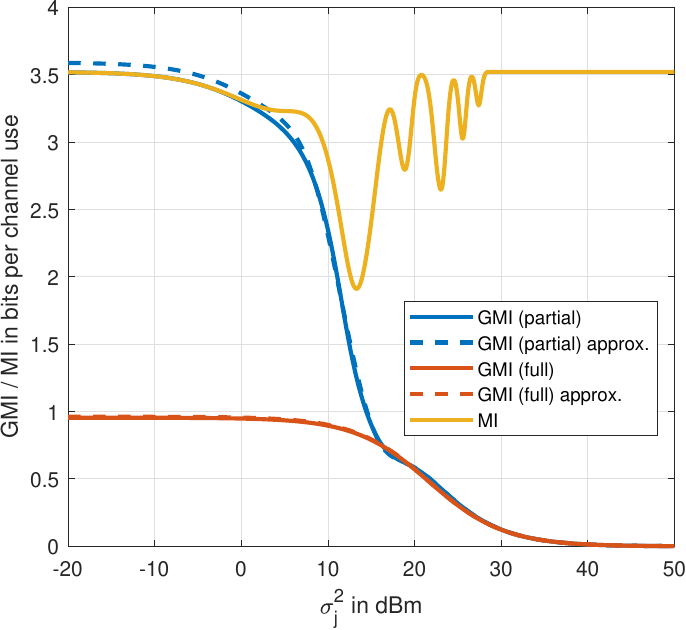}
    \caption{GMI and MI with $\sigma_x^2=20$dBm, $\sigma_i^2=20$dBm and $\sigma_z^2=0$dBm.}
    \label{Fig: GMI_vs_MI_varying_J_pow}
\end{figure}

We also consider a case with fixed $\sigma_x^2$ and $\sigma_i^2$, and with varying $\sigma_j^2$, which leads to Figure. \ref{Fig: GMI_vs_MI_varying_J_pow}. It can be observed that, with low $\sigma_j^2$, there is a gap between ``GMI (partial)'' and ``GMI (full)'' due to the difference in treating $I$ from the corresponding mismatched decoders, while ``GMI (partial)'' is very close to ``MI'' because $q(x,\;y)$ given by (\ref{mismatched decoding metric partial}) is almost $q^\star(x,\;y)$ when $\sigma_j^2$ is negligible. As $\sigma_j^2$ grows, the gap between ``GMI (partial)'' and ``GMI (full)'' shrinks and eventually vanishes. This is because, when interference is dominated by $J$, optimally treating $I$ in decoding leads to limited benefits. As $\sigma_j^2\rightarrow+\infty$, both ``GMI (partial)'' and ``GMI (full)'' tend to be zero, whereas ``MI'' saturates at a high level as stated by Proposition \ref{prop: MI_GMI_saturation} with interchanged $I$ and $J$. Figure \ref{Fig: GMI_vs_MI_varying_J_pow} indicates that treating $j$ as a Gaussian RV is nearly optimal when $\sigma_j^2$ is small, but clearly suboptimal with large $\sigma_j^2$. {In Proposition 5, we formally show that the GMI indeed vanishes as $J$ grows.}

\begin{proposition}\label{prop: GMI_vanish}
With $q(x,\;y)$ given by (\ref{mismatched decoding metric partial}),
\begin{equation}
    \lim_{\sigma_j^2 \rightarrow +\infty}  I_{\text{GMI}}(\mathcal{X},\;\mathcal{I},\;\mathcal{J},\;\sigma_z^2) = 0.
\end{equation}
\end{proposition}
\begin{IEEEproof}
    See Appendix \ref{Appendix: Proof of GMI_vanish}.
\end{IEEEproof}

Another interesting observation from Figure \ref{Fig: GMI_vs_MI_varying_J_pow} is that ``GMI (partial)'' underperforms ``GMI (full)'' in the vicinity of  $\sigma_j^2=17.5$dBm, indicating that the mismatched decoder given by (\ref{mismatched decoding metric partial}) does not always outperform the one given by (\ref{mismatched decoding metric full}), though very often.
Figures \ref{Fig: GMI_vs_MI_varying_X_pow}-\ref{Fig: GMI_vs_MI_varying_J_pow} also depict that (\ref{mismatched capacity approx partial}) and (\ref{mismatched capacity approx full}) provide an adequate approximation of (\ref{mismatched capacity}) under $q(x,\;y)$ given by (\ref{mismatched decoding metric partial}) and (\ref{mismatched decoding metric full}).

\section{Improvement on Decoding metric}
In the spirit of decoding under finite alphabet interference with reduced computational complexity in comparison to maximum-likelihood decoders, we propose two decoding matrices and evaluate their performance.

\subsection{Generalized Gaussian Decoding Metric}
\begin{figure}
    \centering
    \includegraphics[width=6.6cm]{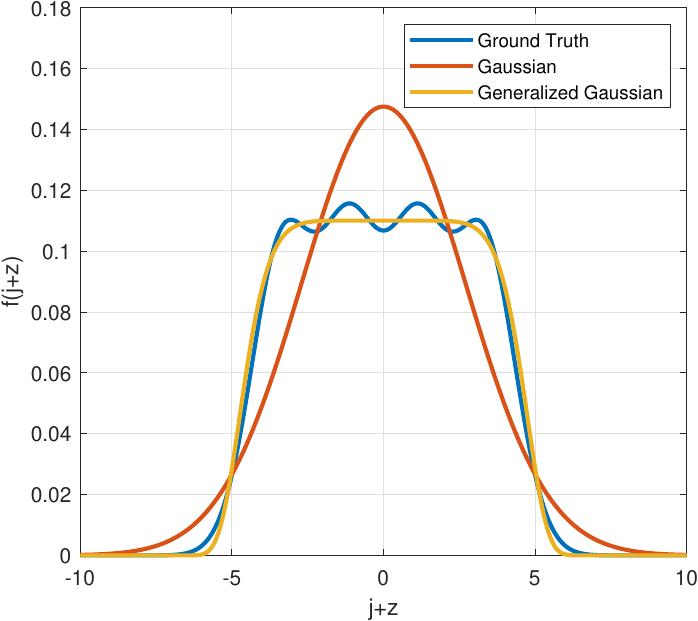}
    \caption{Probability density comparison of ground truth interference plus noise distribution by a mixture of Gaussian, approximation by the Gaussian distribution and Generalized Gaussian distribution with $\beta=5.1$.}
    \label{G_Gauss_illustration}
\end{figure}

In light of improving the similarity between the ground truth interference distribution and the one utilized by the decoding metric, we propose a decoding metric assuming that the real and imaginary parts of $J + Z$ follow i.i.d. generalized Gaussian distributions with their total variance equal to $\sigma_j^2+\sigma_z^2$.\footnote{We consider both real and imaginary part following the same generalize Gaussian distribution. This is sensible if $j$, or a phase-shifted version of $J$, has symmetric distributions in real and imaginary parts, for example, QAMs. For asymmetric distributed $j$, it is straightforward to extend (\ref{G_Gaussian decoding metric}) by assuming asymmetric generalized Gaussian distributions in real and imaginary parts. {For example, in Figure \ref{G_Gauss_illustration}, only the real part is considered to approximate the 4-PAM signal.}} This leads to
\begin{equation}\label{G_Gaussian decoding metric}
    q(x,\;y) = \sum_{\Bar{i}\in\mathcal{I}} \exp\left({-\frac{|\mathfrak{R}\{y-\Bar{i}-x\}|^\beta+|\mathfrak{I}\{y-\Bar{i}-x\}|^\beta}{\alpha^\beta}}\right),
\end{equation}
where $\beta > 0$ is the shape parameter to be chosen, $\alpha$ is the scale parameter and can be determined based on the variance as $\alpha = \sqrt{\frac{(\sigma_j^2+\sigma_z^2)\Gamma(1/\beta)}{2\Gamma(3/\beta)}}$, with $\Gamma(\cdot)$ representing the Gamma function.\footnote{{The normalization constant of the} generalized Gaussian distribution has been ignored in the decoding metric as they do not affect decoding results.}

Figure \ref{G_Gauss_illustration} provides a graphical comparison among a possible ground truth interference plus noise distribution, an
approximation given by a Gaussian distribution whose variance is equal to $\sigma_j^2 + \sigma_z^2$, and an approximation given by a generalized Gaussian distribution with the same variance and $\beta=5.1$. For ease of illustration, the signals are restricted to being real. For clarity,
$\mathcal{I}=\{0\}$, $j$ is drawn from uniform 4-PAM with variance of 8dBm, and $\sigma_z^2=0$dBm . For the generalized Gaussian distribution, $\beta=5.1$ is determined by optimizing the GMI under the generalized Gaussian decoding metric.\footnote{{In the simulations, $\beta$ is determined by exhaustive search to maximize the GMI. In practice, a lookup table for $\beta$ may also be constructed.}} It can be observed that, compared to the Gaussian distribution, the generalized Gaussian distribution provides a better approximation of the ground truth distribution, indicating the potential of utilizing generalized Gaussian distribution in decoding.

\subsection{Interference Decomposition Decoding Metric}
Some constellations can be decomposed into the Minkowski sum of constellations of smaller orders. We propose an interference decomposition decoding metric, which treats one summand constellation optimally and the rest as Gaussian-distributed RVs. Formally, if there exists one decomposition of $\mathcal{J}$ as the following,
\begin{equation}
    \mathcal{J} = \mathcal{J}^+ + \mathcal{J}^-,
\end{equation}
one interference decomposition decoding metric can be
\begin{equation}\label{equ: int_decomp decoding metric}
    q(x,\;y) = \sum_{\Bar{i}\in\mathcal{I}}\sum_{j^+\in\mathcal{J}^+} \exp\left({-\frac{|y-\Bar{i}-j^+-x|^2}{\sigma_{j^-}^2+\sigma_z^2}}\right).
\end{equation}
Note that $|\mathcal{J}^+|<|\mathcal{J}|$ if $|\mathcal{J}^-|\geq2$, therefore, the computational complexity of (\ref{equ: int_decomp decoding metric}) is always lower than (\ref{matched decoding metric}).

\begin{figure}
    \centering
    \includegraphics[width=6.6cm]{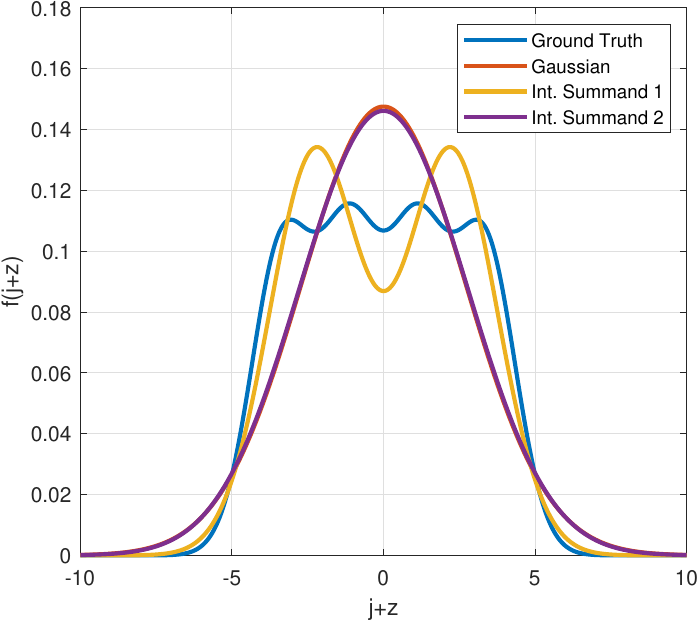}
    \caption{Probability density comparison of ground truth interference plus noise distribution by a mixture of Gaussian, an approximation by Gaussian distribution and two approximations by interference decomposition.}
    \label{Int_decomp_illustration}
\end{figure}
Figure \ref{Int_decomp_illustration} provides a graphical comparison among a
possible ground truth interference plus noise distribution,
a Gaussian approximation, and two approximations given by interference decomposition. The assumptions for interference and noise are the same as in Figure \ref{G_Gauss_illustration}. For interference decomposition, a uniform 4-PAM can be decomposed into two BPSK constellations, where the powers of the two summand constellations are 0.8 and 0.2 of the power of the 4-PAM respectively. By optimally treating either one of the two summand constellations, two different underlying assumptions on interference-plus-noise distributions are obtained and depicted in Figure \ref{Int_decomp_illustration}, where ``Int. Summand 1'' representing $\mathcal{J^+}$ being a BPSK with $\sigma_{j^+}^2=0.8\sigma_j^2$, and ``Int. Summand 2'' representing $\sigma_{j^+}^2=0.2\sigma_j^2$. It can be observed that optimally treating the stronger summand leads to improved approximation on the interference plus noise distribution in comparison with Gaussian approximation, while optimally treating the weaker summand leads to minor changes in the resulting distribution.

\subsection{Results on GMI of the Proposed Decoding Metrics}
\begin{figure}
    \centering
    \includegraphics[width=6.5cm]{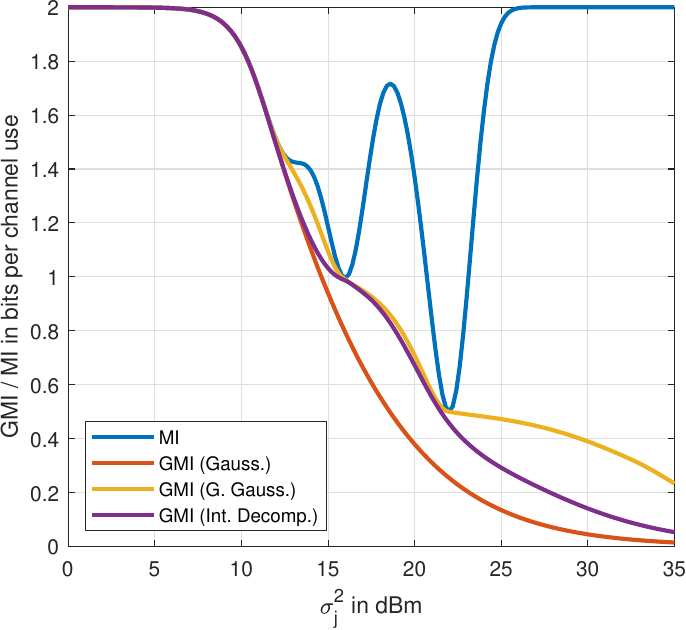}
    \caption{GMI and MI with $\sigma_x^2=15$dBm and $\sigma_z^2=0$dBm.}
    \label{Fig: GMI/MI G Gauss int decomp}
\end{figure}

Consider $x$ and $j$ are generated by QPSK and 16QAM respectively, whereas $\mathcal{I} = \{0\}$ for clarity. From Figure \ref{Fig: GMI/MI G Gauss int decomp}, it can be observed that both the generalized Gaussian decoding metric and the interference decomposition decoding metric lead to improved achievable rates compared to treating interference as a Gaussian RV. Often the generalized Gaussian decoding metric outperforms the interference decomposition decoding metric. However, such superiority comes with an additional computational cost in determining the optimal shape parameter, $\beta$. 
 
\section{Practical Receivers}
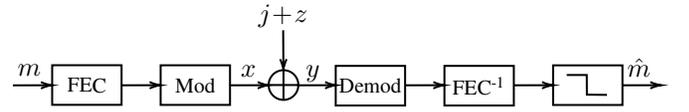
\begin{figure}
    \centering        

\tikzset{every picture/.style={line width=0.75pt}} 

\begin{tikzpicture}[x=0.75pt,y=0.75pt,yscale=-0.5,xscale=0.5]

\draw   (40,70.33) -- (110,70.33) -- (110,110.33) -- (40,110.33) -- cycle ;
\draw   (149.4,70.4) -- (219.4,70.4) -- (219.4,110.4) -- (149.4,110.4) -- cycle ;
\draw   (326,71) -- (396,71) -- (396,111) -- (326,111) -- cycle ;
\draw   (435.67,71) -- (505.67,71) -- (505.67,111) -- (435.67,111) -- cycle ;
\draw   (545.27,70.47) -- (615.27,70.47) -- (615.27,110.47) -- (545.27,110.47) -- cycle ;
\draw    (559.02,80.22) -- (579.27,80.22) -- (579.35,100.05) -- (599.6,100.3) ;
\draw[->]    (110,90.5) -- (149.4,90.5) ;
\draw[->]    (396,90.5) -- (435.67,90.5) ;
\draw[->]    (505.67,90.5) -- (545.27,90.5) ;
\draw[->]    (0.38,90.5) -- (40,90.5) ;
\draw[->]    (615.27,90.5) -- (652.25,90.5) ;
\draw[->]    (219.4,90.5) -- (258.33,90.5) ;
\draw[->]    (288.33,90.5) -- (326,90.5) ;
\draw   (258.33,90.5) .. controls (258.33,82.05) and (265.05,75.33) .. (273.33,75.33) .. controls (281.62,75.33) and (288.33,82.05) .. (288.33,90.5) .. controls (288.33,98.62) and (281.62,105.33) .. (273.33,105.33) .. controls (265.05,105.33) and (258.33,98.62) .. (258.33,90.5) -- cycle ; \draw   (258.33,90.5) -- (288.33,90.5) ; \draw   (273.33,75.33) -- (273.33,105.33) ;
\draw[->]    (273.33,35) -- (273.33,75.33) ;

\draw (75,90.33) node  [rotate=-0.25] [align=left] {\footnotesize FEC};
\draw (184.4,90.4) node  [rotate=-0.25] [align=left] {\footnotesize Mod};
\draw (361,91) node  [rotate=-0.25] [align=left] {\footnotesize Demod};
\draw (470.67,91) node  [rotate=-0.25] [align=left] {\footnotesize FEC\textsuperscript{\mbox{-}1}};
\draw (273,35) node [anchor=south] [inner sep=0.75pt]   [align=left] {$\displaystyle j+z$};
\draw (17,84) node [anchor=south] [inner sep=0.75pt]   [align=left] {$\displaystyle m$};
\draw (238,85) node [anchor=south] [inner sep=0.75pt]   [align=left] {$\displaystyle x$};
\draw (303,90) node [anchor=south] [inner sep=0.75pt]   [align=left] {$\displaystyle y$};
\draw (632,85) node [anchor=south] [inner sep=0.75pt]   [align=left] {$\displaystyle \hat{m}$};
\end{tikzpicture}
    \caption{Transceiver structure.}
    \label{Transceiver structure}
\end{figure}

To shine more light on the designs of the physical layer receivers, the decoding metrics discussed above can be converted into demodulators in BICM-type systems. Consider the physical layer transceiver architecture shown in Figure \ref{Transceiver structure}, where, for clarity, only one interference source, $j$, is considered. In Figure \ref{Transceiver structure}, ``FEC'', ``Mod'', ``Demod'' and ``FEC$^{-1}$'' denote channel encoder,\footnote{This potentially includes interleaving and scrambling.} modulator, demodulator and channel decoder, respectively. The demodulator takes the received signal, $y$, as input and estimates the posterior  probability distribution of the channel input signals, {$\hat{P}_{X\mid Y}(x\mid y)$}, $x\in\mathcal{X}$, based on its knowledge of the input signal, interference and noise. The estimated input distribution can then be converted to log-likelihood-ratios of coded bits, which feed the channel decoder. The decoding metrics mentioned above for the FAGCI model make different assumptions on the interference distribution. The same principle can be applied to the design of demodulators. 

The optimal decoding metric (\ref{matched decoding metric}), which applies the true interference distribution, corresponds to the optimal demodulator, which jointly demodulates $x$ and $j$ as follows.
\begin{equation}\label{Equ: opt demod}
    \hat{P}_{X\mid Y}(x\mid y) = \frac{1}{2\pi\sigma_z^2|\mathcal{J}|}\sum_{\Bar{j}\in\mathcal{J}}\exp\left({-\frac{|y-\Bar{j}-x|^2}{\sigma_z^2}}\right)
\end{equation}
The suboptimal decoding metric (\ref{mismatched decoding metric full}), which treats interference as Gaussian RVs, corresponds to a demodulator which treats interference as Gaussian noise as follows.
\begin{equation}\label{Equ: Gauss demod}
    \hat{P}_{X\mid Y}(x\mid y) = \frac{1}{2\pi(\sigma_j^2+\sigma_z^2)}\exp\left({-\frac{|y-x|^2}{\sigma_j^2+\sigma_z^2}}\right)
\end{equation}
The generalized Gaussian decoding metric given in (\ref{G_Gaussian decoding metric}) can be linked to a demodulator that treats the sum of interference and noise as a generalized-Gaussian-distributed RV, which can be written as 
\begin{equation}\label{Equ: G Gauss demod}
\begin{split}
    &\hat{P}_{X\mid Y}(x\mid y)\\
    &= \left(\frac{\beta}{2\alpha\Gamma(1/\beta)}\right)^2\exp\left({-\frac{|\mathfrak{R}\{y-x\}|^\beta+|\mathfrak{I}\{y-x\}|^\beta}{\alpha^\beta}}\right),
\end{split}
\end{equation}
with $\alpha = \sqrt{\frac{(\sigma_j^2+\sigma_z^2)\Gamma(1/\beta)}{2\Gamma(3/\beta)}}$. Finally, inspired by the interference decomposition decoder in (\ref{equ: int_decomp decoding metric}), an interference decomposition demodulator can be designed to jointly demodulate the input signal and a summand of the interference constellation, while treating the other summand of the interference as a Gaussian RV. This can be written as
\begin{equation}\label{Equ: Int decomp demod}
    \hat{P}_{X\mid Y}(x\mid y) = \frac{1}{2\pi(\sigma_{j^-}^2+\sigma_z^2)|\mathcal{J}^+|}\sum_{j^+\in\mathcal{J}^+} \exp\left({-\frac{|y-j^+-x|^2}{\sigma_{j^-}^2+\sigma_z^2}}\right).
\end{equation}

The similarity between the channel model in Section \ref{Mismatched Decoding for Finite Alphabets} and Figure \ref{Transceiver structure} also implies that the derived GMI in Section \ref{Mismatched Decoding for Finite Alphabets} may resemble the throughput of practical transceivers with the corresponding demodulator.

\begin{figure*}
    \centering
    \includegraphics[width=0.85\linewidth]{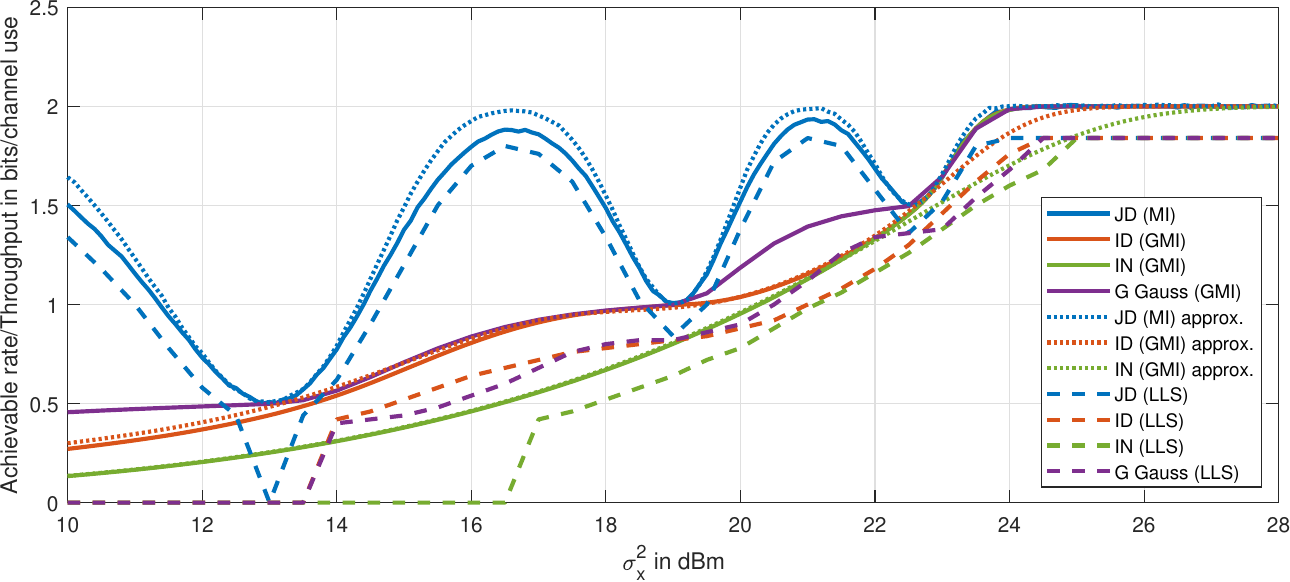}
    \caption{Comparison of GMI/MI and LLS results with $\mathcal{X}$ being QPSK, $\mathcal{J}$ being 16QAM, $\sigma_j^2=20$dBm, $\sigma_z^2$ = 0dBm.}
    \label{Fig: GMI_LLS_QPSK}
    
    \bigskip
    \centering
    \includegraphics[width=0.85\linewidth]{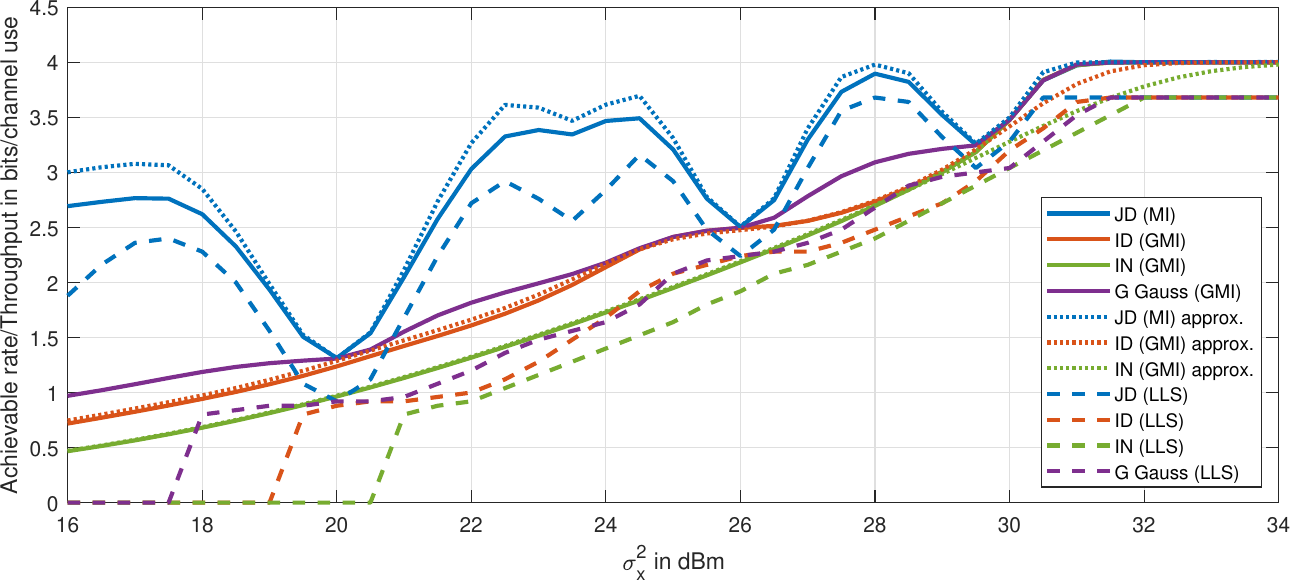}
    \caption{Comparison of GMI/MI and LLS results with $\mathcal{X}$ being 16QAM, $\mathcal{J}$ being 16QAM, $\sigma_j^2=20$dBm, $\sigma_z^2$ = 0dBm.}
    \label{Fig: GMI_LLS_16QAM}
\end{figure*}

We examine the throughput of different decoding metrics when they are used as demodulators and compare it with the GMI led by those decoding metrics in Figures {\ref{Fig: GMI_LLS_QPSK} and \ref{Fig: GMI_LLS_16QAM}}. The legends are explained as follows:
\begin{itemize}
    \item ``JD'' (MI) and (LLS): the optimal decoding, i.e., (\ref{matched decoding metric}) evaluated by MI, and the demodulator jointly demodulating $x$ and $j$, i.e., (\ref{Equ: opt demod}), evaluated by LLS.
    \item ``ID'' (GMI) and (LLS): the interference decomposition decoder, i.e., (\ref{equ: int_decomp decoding metric}) evaluated by GMI, and the interference decomposition demodulator, i.e., (\ref{Equ: Int decomp demod}) evaluated by LLS.
    \item ``IN'' (GMI) and (LLS): the decoder that treats interference as Gaussian RV, i.e., (\ref{mismatched decoding metric full}), evaluated by GMI, and the demodulator that treats interference as Gaussian RV, i.e., (\ref{Equ: Gauss demod}), evaluated by LLS.
    \item ``G Gauss'' (GMI) and (LLS): the generalized Gaussian decoding metric, i.e., (\ref{G_Gaussian decoding metric}), evaluated by GMI, and the generalized Gaussian demodulator, i.e., (\ref{Equ: G Gauss demod}), evaluated by LLS.
\end{itemize}
The channel code used in LLS is the LDPC code specified in \cite{3gpp38212},\footnote{\cite{3gpp38212} specifies the LDPC code for code rate between 0.2 and 0.92. Hence, in Figure \ref{Fig: GMI_LLS_QPSK} and \ref{Fig: GMI_LLS_16QAM}, the code rate suddenly increases from 0 to 0.2 with medium $\sigma_x^2$, and is capped at 0.92 at high $\sigma_x^2$.} with the block length being {4000} symbols. The throughput is determined by searching for the highest code rate that leads to a {block error rate lower than $10^{-2}$}. In Figures \ref{Fig: GMI_LLS_QPSK} and \ref{Fig: GMI_LLS_16QAM}, it can be observed that the proposed demodulators, i.e., (\ref{Equ: G Gauss demod}) and (\ref{Equ: Int decomp demod}), lead to improved throughput compared to that given by (\ref{Equ: Gauss demod}). In addition, GMI/MI provides adequate prediction on the physical layer throughput when the demodulator used in the LLS corresponds to the decoding metric assumed by GMI/MI. The gap between GMI/MI and throughput is due to the non-ideal encoding and decoding.

These results lead to two key implications:
\begin{enumerate}
    \item GMI/MI provides alternative tools to LLS for predicting demodulator performance.\footnote{Note that another approach for demodulator evaluation is proposed in \cite{Fertl_TSP}, where interference is not considered, and the resulting expression is not in closed form, hence not viable for system design.} The computational complexity for computing GMI/MI is much lower than that of running LLS, making the former promising. The approximations by GMI/MI also provide adequate prediction with even less computational burden.
    \item GMI/MI can be important for accurate physical layer abstraction in SLS. Optimizing GMI/MI indirectly optimizes the physical layer throughput under the corresponding demodulators.
\end{enumerate}

The second implication motivates the next section, where an example on GMI optimization for MU-MISO systems is explored.

\section{Signal Processing with GMI}
In multi-user communication systems,  it is common that multi-user interference is not optimally treated due to reasons such as limited receiver complexity and lack of knowledge on interference. The channel model, the mismatched decoder, the corresponding GMI and its approximation from Section \ref{Mismatched Decoding for Finite Alphabets} provide novel tools for analyzing such scenarios and designing signal processing methods. In this section, we extend the results from the SISO FAGCI model to multi-user MISO systems to enable novel signal processing algorithms. In particular, we provide an example on precoder optimization with the objective of maximizing sum-rates under different decoding strategies adapted by users.

\begin{figure}
    \centering
\tikzset{every picture/.style={line width=0.75pt}} 

\begin{tikzpicture}[x=0.75pt,y=0.75pt,yscale=-0.9,xscale=0.9]

\draw   (169.4,62.29) -- (219.4,62.29) -- (219.4,145.42) -- (169.4,145.42) -- cycle ;
\draw    (219.4,70.43) -- (247.17,70.43) ;
\draw    (247.17,70.43) -- (247.17,61.13) ;
\draw   (247.17,61.12) -- (237.85,52.78) -- (256.52,52.81) -- cycle ;
\draw    (219.4,93.68) -- (247.17,93.68) ;
\draw    (247.17,93.68) -- (247.17,84.38) ;
\draw   (247.17,84.38) -- (237.85,76.03) -- (256.52,76.07) -- cycle ;
\draw    (219.4,116.74) -- (247.17,116.74) ;
\draw    (247.17,116.74) -- (247.17,107.44) ;
\draw   (247.17,107.44) -- (237.85,99.09) -- (256.52,99.13) -- cycle ;
\draw    (219.63,139.99) -- (247.4,139.99) ;
\draw    (247.4,139.99) -- (247.4,130.69) ;
\draw   (247.4,130.69) -- (238.09,122.34) -- (256.76,122.38) -- cycle ;
\draw   (413.97,39.73) -- (479.87,39.73) -- (479.87,70.74) -- (413.97,70.74) -- cycle ;
\draw    (386.39,55.08) -- (414.16,55.08) ;
\draw    (386.86,55.08) -- (386.86,45.78) ;
\draw   (386.86,45.78) -- (377.55,37.43) -- (396.22,37.47) -- cycle ;
\draw   (414.1,91.6) -- (480,91.6) -- (480,122.6) -- (414.1,122.6) -- cycle ;
\draw    (386.15,106.72) -- (413.92,106.72) ;
\draw    (386.62,106.72) -- (386.62,97.42) ;
\draw   (386.62,97.41) -- (377.3,89.07) -- (395.97,89.1) -- cycle ;
\draw   (413.97,144.74) -- (479.87,144.74) -- (479.87,175.74) -- (413.97,175.74) -- cycle ;
\draw    (386.12,159.94) -- (413.89,159.94) ;
\draw    (386.59,159.94) -- (386.59,150.64) ;
\draw   (386.59,150.64) -- (377.28,142.3) -- (395.95,142.33) -- cycle ;
\draw [color={rgb, 255:red, 243; green, 177; blue, 32 }  ,draw opacity=1 ] [dash pattern={on 4.5pt off 4.5pt}]  (263.36,85.46) -- (365.4,48.63) ;
\draw [shift={(367.28,47.95)}, rotate = 160.15] [color={rgb, 255:red, 243; green, 177; blue, 32 }  ,draw opacity=1 ][line width=0.75]    (6.56,-1.97) .. controls (4.17,-0.84) and (1.99,-0.18) .. (0,0) .. controls (1.99,0.18) and (4.17,0.84) .. (6.56,1.97)   ;
\draw [color={rgb, 255:red, 243; green, 177; blue, 32 }  ,draw opacity=1 ]   (263.36,97.24) -- (368.67,96.94) ;
\draw [shift={(370.67,96.93)}, rotate = 179.83] [color={rgb, 255:red, 243; green, 177; blue, 32 }  ,draw opacity=1 ][line width=0.75]    (6.56,-1.97) .. controls (4.17,-0.84) and (1.99,-0.18) .. (0,0) .. controls (1.99,0.18) and (4.17,0.84) .. (6.56,1.97)   ;
\draw [color={rgb, 255:red, 243; green, 177; blue, 32 }  ,draw opacity=1 ] [dash pattern={on 4.5pt off 4.5pt}]  (262.79,109.8) -- (364.09,147.55) ;
\draw [shift={(365.97,148.25)}, rotate = 200.44] [color={rgb, 255:red, 243; green, 177; blue, 32 }  ,draw opacity=1 ][line width=0.75]    (6.56,-1.97) .. controls (4.17,-0.84) and (1.99,-0.18) .. (0,0) .. controls (1.99,0.18) and (4.17,0.84) .. (6.56,1.97)   ;
\draw [color={rgb, 255:red, 208; green, 2; blue, 27 }  ,draw opacity=1 ]   (260.91,81.28) -- (362.95,44.44) ;
\draw [shift={(364.84,43.76)}, rotate = 160.15] [color={rgb, 255:red, 208; green, 2; blue, 27 }  ,draw opacity=1 ][line width=0.75]    (6.56,-1.97) .. controls (4.17,-0.84) and (1.99,-0.18) .. (0,0) .. controls (1.99,0.18) and (4.17,0.84) .. (6.56,1.97)   ;
\draw [color={rgb, 255:red, 74; green, 144; blue, 226 }  ,draw opacity=1 ] [dash pattern={on 4.5pt off 4.5pt}]  (265.8,89.65) -- (367.85,52.81) ;
\draw [shift={(369.73,52.13)}, rotate = 160.15] [color={rgb, 255:red, 74; green, 144; blue, 226 }  ,draw opacity=1 ][line width=0.75]    (6.56,-1.97) .. controls (4.17,-0.84) and (1.99,-0.18) .. (0,0) .. controls (1.99,0.18) and (4.17,0.84) .. (6.56,1.97)   ;
\draw [color={rgb, 255:red, 208; green, 2; blue, 27 }  ,draw opacity=1 ] [dash pattern={on 4.5pt off 4.5pt}]  (262.98,92.28) -- (368.3,91.98) ;
\draw [shift={(370.3,91.97)}, rotate = 179.83] [color={rgb, 255:red, 208; green, 2; blue, 27 }  ,draw opacity=1 ][line width=0.75]    (6.56,-1.97) .. controls (4.17,-0.84) and (1.99,-0.18) .. (0,0) .. controls (1.99,0.18) and (4.17,0.84) .. (6.56,1.97)   ;
\draw [color={rgb, 255:red, 74; green, 144; blue, 226 }  ,draw opacity=1 ] [dash pattern={on 4.5pt off 4.5pt}]  (263.73,102.21) -- (369.05,101.9) ;
\draw [shift={(371.05,101.9)}, rotate = 179.83] [color={rgb, 255:red, 74; green, 144; blue, 226 }  ,draw opacity=1 ][line width=0.75]    (6.56,-1.97) .. controls (4.17,-0.84) and (1.99,-0.18) .. (0,0) .. controls (1.99,0.18) and (4.17,0.84) .. (6.56,1.97)   ;
\draw [color={rgb, 255:red, 208; green, 2; blue, 27 }  ,draw opacity=1 ] [dash pattern={on 4.5pt off 4.5pt}]  (265.24,105.31) -- (366.54,143.05) ;
\draw [shift={(368.41,143.75)}, rotate = 200.44] [color={rgb, 255:red, 208; green, 2; blue, 27 }  ,draw opacity=1 ][line width=0.75]    (6.56,-1.97) .. controls (4.17,-0.84) and (1.99,-0.18) .. (0,0) .. controls (1.99,0.18) and (4.17,0.84) .. (6.56,1.97)   ;
\draw [color={rgb, 255:red, 74; green, 144; blue, 226 }  ,draw opacity=1 ]   (260.72,113.68) -- (362.02,151.42) ;
\draw [shift={(363.89,152.12)}, rotate = 200.44] [color={rgb, 255:red, 74; green, 144; blue, 226 }  ,draw opacity=1 ][line width=0.75]    (6.56,-1.97) .. controls (4.17,-0.84) and (1.99,-0.18) .. (0,0) .. controls (1.99,0.18) and (4.17,0.84) .. (6.56,1.97)   ;

\draw (186,95) node [anchor=north west][inner sep=0.75pt]   [align=left] {Tx};
\draw (426,48) node [anchor=north west][inner sep=0.75pt]  [color={rgb, 255:red, 208; green, 2; blue, 27 }  ,opacity=1 ] [align=left] {User-1};
\draw (426,100) node [anchor=north west][inner sep=0.75pt]  [color={rgb, 255:red, 243; green, 177; blue, 32 }  ,opacity=1 ] [align=left] {User-2};
\draw (426,153) node [anchor=north west][inner sep=0.75pt]  [color={rgb, 255:red, 74; green, 144; blue, 226 }  ,opacity=1 ] [align=left] {User-3};
\end{tikzpicture}

\caption{{Illustration of a MISO broadcast channel, where solid arrows denote desired signals and dashed arrows denote interfering signals. The three users potentially suffer from multi-user interference.}}
\label{Fig: MU-MISO illustration}
\end{figure}

\subsection{Achievable Rate of MU-MISO Incorporating Decoding Strategy}
Consider a downlink MU-MISO system with $N_{\text{T}}$ antennas at the transmitter and $K$ single-antenna users {as illustrated in Figure \ref{Fig: MU-MISO illustration}}, where the transmitted signal can be written as
\begin{equation}\label{Equ: single-user MISO Tx}
    \mathbf{x} = \mathbf{P}\mathbf{s},
\end{equation}
where $\mathbf{P} = [\mathbf{p}_1,\; \mathbf{p}_2,\;...,\;\mathbf{p}_K]\in\mathcal{C}^{N_\text{T}\times K}$ is the precoding matrix, and $\mathbf{s} = [s_1,\;s_2,\;...,\;s_K]^T \in\mathcal{C}^{K\times 1}$ is the symbol vector to be transmitted. User-$k$ desires $s_k$ and generally suffers from multi-user interference and noise. For simplicity, we assume that all users adopt the same constellation, i.e., $s_k\in\mathcal{X}$, $\forall k \in \{1,\;...,\;k\}$. 

Given a decoding strategy, i.e., deciding which interfering streams to be treated optimally and as Gaussian RV at each user, the achievable rate at each user is given by extending the GMI in (\ref{mismatched capacity}). Consider a generic expression for the received signal at user-$k$, $\forall k \in \{1,\;...,\;K\}$, as follows: 
\begin{equation}\label{Equ: single-user MISO Rx}
    y_k = \mathbf{h}_k^H(\mathbf{P}_x\mathbf{x} + \mathbf{P}_i\mathbf{i} + \mathbf{P}_j\mathbf{j}) + z_k,
\end{equation}
where $\mathbf{h}_k$ is the channel vector at user-$k$, and $\mathbf{x}$, $\mathbf{i}$ and $\mathbf{j}$ are respectively the desired symbol vector for user-$k$, interfering symbol vector to be treated optimally and interfering symbol vector to be treated as Gaussian RVs by user-$k$, respectively drawn from $\mathcal{X}$, $\mathcal{I}$ and $\mathcal{J}$. $\mathbf{P}_x$, $\mathbf{P}_i$ and $\mathbf{P}_j$ are their precoders. $z_k$ is additive noise generated from i.i.d. $\mathcal{CN}(0,\sigma_z^2)$. Without loss of generality, we assume that $\mathbb{E}[\mathbf{x} \mathbf{x}^H] = \mathbf{I}^{\text{dim}(\mathbf{x})}$, $\mathbb{E}[\mathbf{i} \mathbf{i}^H] = \mathbf{I}^{\text{dim}(\mathbf{i})}$ and $\mathbb{E}[\mathbf{j} \mathbf{j}^H] = \mathbf{I}^{\text{dim}(\mathbf{j})}$ unless any of the constellations is $\{0\}$. WLOG, user-$k$ utilizes a decoding metric given by
\begin{equation}
    q(\mathbf{x},\;y_k) = \sum_{\Bar{\mathbf{i}}\in\mathcal{I}} \exp\left({-\frac{|y_k-\mathbf{h}_k^H\mathbf{P}_i\Bar{\mathbf{i}}-\mathbf{h}_k^H\mathbf{P}_x\mathbf{x}|^2}{\|\mathbf{h}_k^H\mathbf{P}_j\|^2 +\sigma_z^2}}\right).
\end{equation}

\begin{figure*}
\begin{flalign}\label{MISO GMI}
    &I_{\text{GMI},k}\left(\mathbf{h}_k^H\mathbf{P}_{x}\mathcal{X},\; \mathbf{h}_k^H\mathbf{P}_{i}\mathcal{I},\; \mathbf{h}_k^H\mathbf{P}_{j}\mathcal{J},\; \sigma_z^2\right) &&\nonumber\\
    =&\sup_{s\geq 0}\;\;\log|\mathcal{X}|-\frac{1}    {|\mathcal{X}||\mathcal{I}||\mathcal{J}|} \sum_{\mathbf{x}\in\mathcal{X},\;\mathbf{i}\in\mathcal{I},\;\mathbf{j}\in\mathcal{J}}\mathbb{E}_Z\left[ \log \sum_{\Bar{\mathbf{x}}\in\mathcal{X}} \left( \sum_{\Bar{\mathbf{i}}\in\mathcal{I}} \exp\left(-\frac{|\mathbf{h}_k^H\mathbf{P}_{x}(\mathbf{x}-\Bar{\mathbf{x}})+\mathbf{h}_k^H\mathbf{P}_{\mathbf{I}}(\mathbf{i}-\Bar{\mathbf{i}})+\mathbf{h}_k^H\mathbf{P}_{j}\mathbf{j}+Z|^2}{\|\mathbf{h}_k^H\mathbf{P}_{j}\|^2 + \sigma_z^2}\right) \right)^s\right] &&\nonumber\\
    &+ \frac{s}{|\mathcal{I}||\mathcal{J}|} \sum_{\mathbf{i}\in\mathcal{I},\;\mathbf{j}\in\mathcal{J}} \mathbb{E}_Z\left[ \log \left(\sum_{\Bar{i}\in\mathcal{I}} \exp\left(-\frac{|\mathbf{h}_k^H\mathbf{P}_{i}(\mathbf{i}-\Bar{\mathbf{i}})+\mathbf{h}_k^H\mathbf{P}_{j}\mathbf{j}+Z|^2}{\|\mathbf{h}_k^H\mathbf{P}_{j}\|^2 + \sigma_z^2}\right) \right) \right] &&
\end{flalign}
\begin{flalign}\label{MISO GMI approx}                      &I_{\text{GMI},k}^{\text{approx}}\left(\mathbf{h}_k^H\mathbf{P}_{x} \mathcal{X},\; \mathbf{h}_k^H\mathbf{P}_{i}\mathcal{I},\; \mathbf{h}_k^H\mathbf{P}_{j}\mathcal{J},\; \sigma_z^2\right) &&\nonumber\\ 
    =& \log|\mathcal{X}| - \frac{1}{|\mathcal{X}||\mathcal{I}||\mathcal{J}|} \sum_{\mathbf{x}\in\mathcal{X},\;\mathbf{i}\in\mathcal{I},\;\mathbf{j}\in\mathcal{J}} \log \sum_{\Bar{\mathbf{x}}\in\mathcal{X},\;\Bar{\mathbf{i}}\in\mathcal{I}} \exp\left(-\frac{|\mathbf{h}_k^H\mathbf{P}_{x}(\mathbf{x}-\Bar{\mathbf{x}})+\mathbf{h}_k^H\mathbf{P}_{i}(\mathbf{i}-\Bar{\mathbf{i}})+\mathbf{h}_k^H\mathbf{P}_{j}\mathbf{j}|^2}{\|\mathbf{h}_k^H\mathbf{P}_{j}\|^2+2\sigma_z^2}\right) &&\nonumber\\ 
    & + \frac{1}{|\mathcal{I}||\mathcal{J}|} \sum_{\mathbf{i}\in\mathcal{I},\;\mathbf{j}\in\mathcal{J}} \log \sum_{\Bar{\mathbf{i}}\in\mathcal{I}} \exp\left(-\frac{|\mathbf{h}_k^H\mathbf{P}_{i}(\mathbf{i}-\Bar{\mathbf{i}})+\mathbf{h}_k^H\mathbf{P}_{j}\mathbf{j}|^2}{\|\mathbf{h}_k^H\mathbf{P}_{j}\|^2+2\sigma_z^2}\right) &&
\end{flalign}
\begin{flalign}\label{MISO GMI approx gradient}
    &\nabla_{\widetilde{\mathbf{P}}_k}
    I_{\text{GMI},k}^{\text{approx}}\left(\mathbf{h}_k^H\mathbf{P}_{x} \mathcal{X},\; \mathbf{h}_k^H\mathbf{P}_{i}\mathcal{I},\; \mathbf{h}_k^H\mathbf{P}_{j}\mathcal{J},\; \sigma_z^2\right) &&\nonumber\\
    =& -\frac{1}{|\mathcal{X}||\mathcal{I}||\mathcal{J}|} \sum_{\mathbf{x}\in\mathcal{X},\;\mathbf{i}\in\mathcal{I},\;\mathbf{j}\in\mathcal{J}} \frac{1}{\sum_{\Bar{\mathbf{x}}\in\mathcal{X},\;\Bar{\mathbf{i}}\in\mathcal{I}}g(\widetilde{\mathbf{P}}_k,\mathbf{v})}
    \sum_{\Bar{\mathbf{x}}\in\mathcal{X},\;\Bar{\mathbf{i}}\in\mathcal{I}}g_1(\widetilde{\mathbf{P}}_k,\mathbf{v})
    + \frac{1}{|\mathcal{I}||\mathcal{J}|} \sum_{\mathbf{i}\in\mathcal{I},\;\mathbf{j}\in\mathcal{J}} \frac{1}{\sum_{\Bar{\mathbf{i}}\in\mathcal{I}} h(\widetilde{\mathbf{P}}_k,\mathbf{w})}
    \sum_{\Bar{\mathbf{i}}\in\mathcal{I}} h_1(\widetilde{\mathbf{P}}_k,\mathbf{w}) &&
\end{flalign}
\begin{flalign}
    \;\;\;\;\;\;\; &g(\widetilde{\mathbf{P}}_k,\mathbf{v}) = \exp\left(-\frac{|\mathbf{h}_k^H\widetilde{\mathbf{P}}_k\mathbf{v}|^2}{\|\mathbf{h}_k^H\widetilde{\mathbf{P}}_k\;\mathbf{I}_{j}\|^2 + 2\sigma_z^2}\right) &&\label{g(Pv)}\\ 
    &h(\widetilde{\mathbf{P}}_k,\mathbf{w}) = \exp\left(-\frac{|\mathbf{h}_k^H\widetilde{\mathbf{P}}_k\mathbf{w}|^2}{\|\mathbf{h}_k^H\widetilde{\mathbf{P}}_k\;\mathbf{I}_{j}\|^2 + 2\sigma_z^2}\right) &&\label{h(Pv)}\\
    &g_1(\widetilde{\mathbf{P}}_k,\mathbf{v}) = g(\widetilde{\mathbf{P}}_k,\mathbf{v})\left(\frac{-\left(\|\mathbf{h}_k^H\widetilde{\mathbf{P}}_k\;\mathbf{I}_{j}\|^2 + 2\sigma_z^2\right)\mathbf{h}_k\mathbf{h}_k^H\widetilde{\mathbf{P}}_k\mathbf{v}\mathbf{v}^H + (\mathbf{v}^H\widetilde{\mathbf{P}}_k^H\mathbf{h}_k\mathbf{h}_k^H\widetilde{\mathbf{P}}_k\mathbf{v}) \mathbf{h}_k\mathbf{h}_k^H\widetilde{\mathbf{P}}_k\;\mathbf{I}_{j}\;\mathbf{I}_{j}^H}{\left(\|\mathbf{h}_k^H\widetilde{\mathbf{P}}_k\;\mathbf{I}_{j}\|^2 + 2\sigma_z^2\right)^2}\right) &&\label{g1(Pv)}\\
    &h_1(\widetilde{\mathbf{P}}_k,\mathbf{w}) = h(\widetilde{\mathbf{P}}_k,\mathbf{w})\left(\frac{-\left(\|\mathbf{h}_k^H\widetilde{\mathbf{P}}_k\;\mathbf{I}_{j}\|^2 + 2\sigma_z^2\right)\mathbf{h}_k\mathbf{h}_k^H\widetilde{\mathbf{P}}_k\mathbf{w}\mathbf{w}^H + (\mathbf{w}^H\widetilde{\mathbf{P}}_k^H\mathbf{h}_k\mathbf{h}_k^H\widetilde{\mathbf{P}}_k\mathbf{w}) \mathbf{h}_k\mathbf{h}_k^H\widetilde{\mathbf{P}}_k\;\mathbf{I}_{j}\;\mathbf{I}_{j}^H}{\left(\|\mathbf{h}_k^H\widetilde{\mathbf{P}}_k\;\mathbf{I}_{j}\|^2 + 2\sigma_z^2\right)^2}\right) &&\label{h1(Pv)}
\end{flalign}
\hrule
\end{figure*}

By extending (\ref{mismatched capacity}) and (\ref{mismatched capacity approx partial}), the GMI at user-$k$ can be expressed as (\ref{MISO GMI}) and approximated by (\ref{MISO GMI approx}). To reduce the computational complexity, we focus on (\ref{MISO GMI approx}) during signal processing, but the results will be evaluated by (\ref{MISO GMI}).
To enable optimization of GMI, the gradient of GMI w.r.t. precoders is given by (\ref{MISO GMI approx gradient}), where $g(\widetilde{\mathbf{P}}_k,\mathbf{v})$, $h(\widetilde{\mathbf{P}}_k,\mathbf{v})$, $g_1(\widetilde{\mathbf{P}}_k,\mathbf{v})$ and $h_1(\widetilde{\mathbf{P}}_k,\mathbf{v})$ are given by (\ref{g(Pv)}-\ref{h1(Pv)}), and
\begin{equation}
\widetilde{\mathbf{P}}_k = [\mathbf{P}_x\;\; \mathbf{P}_i\;\; \mathbf{P}_j],
\end{equation}
\begin{equation}
\mathbf{v} = [\mathbf{x}-\Bar{\mathbf{x}};\;\;\;\; \mathbf{i}-\Bar{\mathbf{i}};\;\;\;\; \mathbf{j}],
\end{equation}
\begin{equation}
\mathbf{w} = [\mathbf{0}^{\text{dim}(\mathcal{X})\times 1};\;\;\;\; \mathbf{i}-\Bar{\mathbf{i}};\;\;\;\; \mathbf{j}],
\end{equation}
\begin{equation}
\mathbf{I}_j= [\mathbf{0}^{\text{dim}(\mathcal{X})\times \text{dim}(\mathcal{J})};\;\;\;\; \mathbf{0}^{\text{dim}(\mathcal{I})\times \text{dim}(\mathcal{J})};\;\;\;\; \mathbf{I}^{\text{dim}(\mathcal{J})}].
\end{equation}
Note that $\widetilde{\mathbf{P}}_k$ is a column-wise permutation of the precoding matrix at the transmitter, $\mathbf{P}$, based on user-$k$'s decoding strategy.

\subsection{Sum-Rate Optimization}
We consider a sum-rate maximization problem under fixed transmit power budget as follows:
\begin{equation}
\begin{split}
    \mathcal{P}_1: \;\; \underset{\mathbf{P}}{\text{max}} \: \: & \sum_{k\in\mathcal{K}} I_{\text{GMI},k}^\text{approx}\\ 
    \text{s.t.} \: \: \: \;
    & \|\mathbf{P}\|_\text{F}^2 \leq P_\text{T},
\end{split}
\end{equation}
where $P_\text{T}$ is the transmit power budget. Note that the approximated GMI by (\ref{MISO GMI approx}) is used to reduce computational complexity. The logarithmic barrier \cite{boyd2004convex} is used to approximate $\mathcal{P}_1$ into the following unconstrained optimization problem
\begin{equation}\label{SR_unconstrained_form}
\begin{split}
    \mathcal{P}_2: \;\; \underset{\mathbf{P}}{\text{max}} \;\; 
    & \Omega(\mathbf{P}) \triangleq \tau\:\sum_{k\in\mathcal{K}} I_{\text{GMI},k}^\text{approx} + \log \left(P_\text{T} - \|\mathbf{P}\|_\text{F}^2\right),
\end{split}
\end{equation}
where $\tau$ is a positive parameter that sets the accuracy of the approximation. $\mathcal{P}_2$ can be solved by gradient ascent as given in Algorithm \ref{Algo: GD for SR opt}, where $\nabla_\mathbf{P} I_\text{sum}^\text{approx}(\mathbf{P})$ is computed by adding the GMI gradient at all users given by (\ref{MISO GMI approx gradient}). {The convergence of Algorithm \ref{Algo: GD for SR opt} is ensured by employing a backtracking line search to determine the step sizes.}

\begin{algorithm}[!t]
    \caption{Gradient Ascent with Barrier Method for Sum-Rate Maximization}
    \label{Algo: GD for SR opt}
    \begin{algorithmic}[1]
        \REQUIRE $P_\text{T}$, $\mathcal{X}_k$, $\mathbf{h}_k$, $k\in\mathcal{K}$, $\beta$, $\epsilon$, $v_\text{max}$, $\tau_\text{max}$.
        \ENSURE $\mathbf{P}^\star$
        \STATE Initialize $\mathbf{P}^0$ {randomly and $\tau:=1$}.
        \REPEAT 
            \STATE $v:=0$.
            \REPEAT
                \STATE $\Delta\mathbf{P} := \tau\:\nabla_\mathbf{P} I_\text{sum}^\text{approx}(\mathbf{P}^v) + \frac{\mathbf{P}^v}{\| \mathbf{P}^v \|_\text{F}^2 - P_\text{T}}$.
                \STATE Choose step size $\alpha$ via backtracking line search.
                \STATE Update. $\mathbf{P}^{v+1} := \mathbf{P}^v + \alpha \Delta\mathbf{P}$.
                \STATE Update. $\Omega^{v+1} := \Omega(\mathbf{P}^{v+1})$.
                \STATE $v:=v+1$.
            \UNTIL $\mid \Omega^{v} - \Omega^{v-1} \mid\; < \epsilon$ or $v \geq v_\text{max}$.
            \STATE Update $\tau := \beta \tau$.
        \UNTIL $\tau \geq \tau_\text{max}$.
        \RETURN $\mathbf{P}^\star := \mathbf{P}^{v}$.
    \end{algorithmic}
\end{algorithm}

\subsection{Numerical Simulations}
\subsubsection{Channel Model}
Consider spatially correlated Rayleigh fading channels, i.e., $\mathbf{h}_k \sim \mathcal{CN}(\mathbf{0}_{N_\text{T} \times 1},\;\mathbf{R}_k)$, where $\mathbf{R}_k \in \mathbb{C}^{N_\text{T} \times N_\text{T}}$ is the channel covariance matrix for user-$k$. Applying eigendecomposition, one can have 
\begin{equation}
    \mathbf{R}_k = \mathbf{U}_k \mathbf{\Lambda}_k \mathbf{U}_k^H, 
\end{equation}
where $\mathbf{\Lambda}_k$ is a $r_k \times r_k$ diagonal matrix of nonzero eigenvalues, and $\mathbf{U}_k \in \mathbb{C}^{N_\text{T} \times r_k} $ is a matrix whose columns are the associated eigenvectors of $\mathbf{R}_k$. With the Karhunen-Loeve model, $\mathbf{h}_k$ can be equivalently expressed as
\begin{equation}
    \mathbf{h}_k = \mathbf{U}_k\mathbf{\Lambda}_k^{\frac{1}{2}}\mathbf{w}_k,
\end{equation}
where $\mathbf{w}_k \in \mathbb{C}^{r_k \times 1} \sim \mathcal{CN}(\mathbf{0}^{r_k \times 1},\;\mathbf{I}^{r_k})$.

We consider a geometrical one-ring scattering model \cite{one_ring_model}, where the underlying assumption is that there is no line-of-sight path between the transmitter and user-$k$, and the angle-of-departure (AoD) of all the scatters w.r.t. the transmitter is uniformly distributed around a centre AoD, denoted by $\theta_k$,\footnote{$\theta_k$ measures the angle between x-axis and the centre AoD.} with an angular spread of $\Delta_k$. We assume that the transmitter is equipped with a linear and uniform antenna array centered at the origin and along the $y$-axis with an antenna element spacing of $\lambda/2$. The $(m,n)$-th elements of $\mathbf{R}_k$ is given by
\begin{equation}
    [\mathbf{R}_k]_{m,n} = \frac{1}{2\Delta_k} \int_{\theta_k-\Delta_k}^{\theta_k+\Delta_k}
    e^{-j\pi (m-n) \sin{\alpha}} d\alpha.
\end{equation}

In this work, we model correlated user channels where interference may hinder performance. As a result, we set $\theta_k=\pi/3$ and $\Delta_k=\pi/6$, $\forall k\in \{1,...,K\}$.

\subsubsection{Numerical Results}
\begin{figure}
    \centering
    \includegraphics[width=6.5cm]{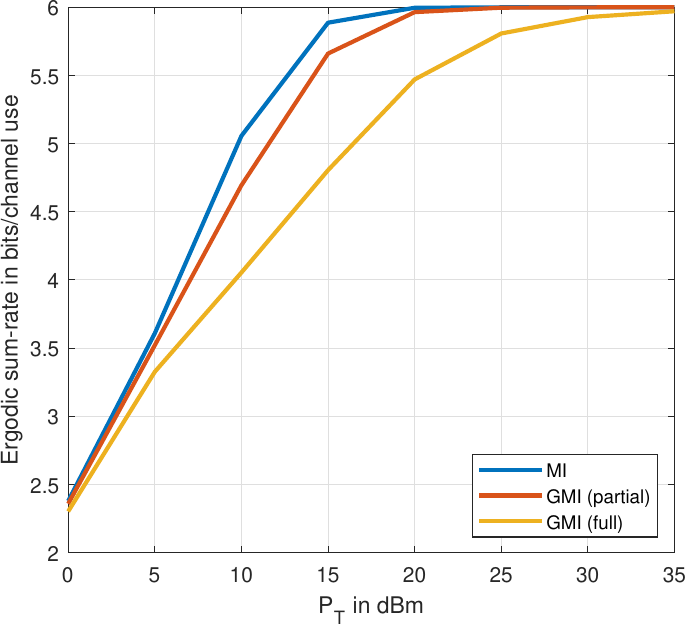}
    \caption{Sum-rate comparison of MU-MISO under different decoding strategies.}
    \label{Fig: MISO_results}
\end{figure}

With $N_\text{T}=4$, $K=3$ and $\mathcal{X}$ given by QPSK, we evaluate the optimal sum-rate of the system under different decoding strategies deployed by the users. The results are depicted in Figure \ref{Fig: MISO_results}, where the legends and their corresponding decoding strategies are as follows.
\begin{enumerate}
    \item ``MI'' indicates that the optimization objective is the mutual information, whose underlying assumption is that all users apply the optimal decoding metric, which treats the other two users' interference optimally;
    \item ``GMI (partial)'' implies optimizing a GMI which assumes that each user treats interference from one of the other users optimally and the interference from the third as Gaussian RV. In particular, we assume that user-1 treats user-2's interference optimally, user-2 treats user-3's interference optimally, and user-3 treats user-1's interference optimally;
    \item ``GMI (full)'' indicates that the objective function is the GMI expression which assumes that each user treats all multi-user interference as Gaussian RV.
\end{enumerate}

\begin{figure}
    \centering
    \includegraphics[width=6.5cm]{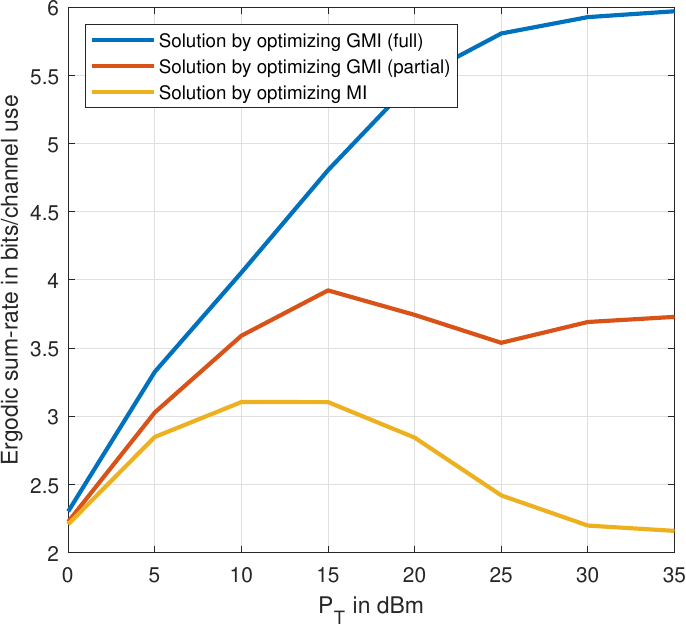}
    \caption{Sum-rate comparison of MU-MISO, with decoders treating interference as Gaussian RV and precoders by optimizing different objective functions.}
    \label{Fig: MISO_precoder_comparison}
\end{figure}

   From Figure \ref{Fig: MISO_results}, it can be observed that different decoding metrics lead to different sum-rates in general. The order is always ``MI'' $>$ ``GMI (partial)'' $>$ ``GMI (full)'', which aligns with the intuition that utilizing more knowledge on interference leads to better decoding, and therefore better performance. Figure \ref{Fig: MISO_precoder_comparison} evaluates the ``GMI (full)'' but with precoders obtained by optimizing different objective functions as in Figure \ref{Fig: MISO_results}. It shows that not only the GMI optimization solution for one decoding metric is different from other decoding metrics, but also the mismatch in objective function can even be harmful, noticing the non-monotonic change in sum-rate under discrepancy between the decoding metric and the objective function.
   This indicates that mismatched receivers require dedicated signal processing approaches based on the corresponding GMI.

\begin{figure}
    \centering
    \includegraphics[width=6.5cm]{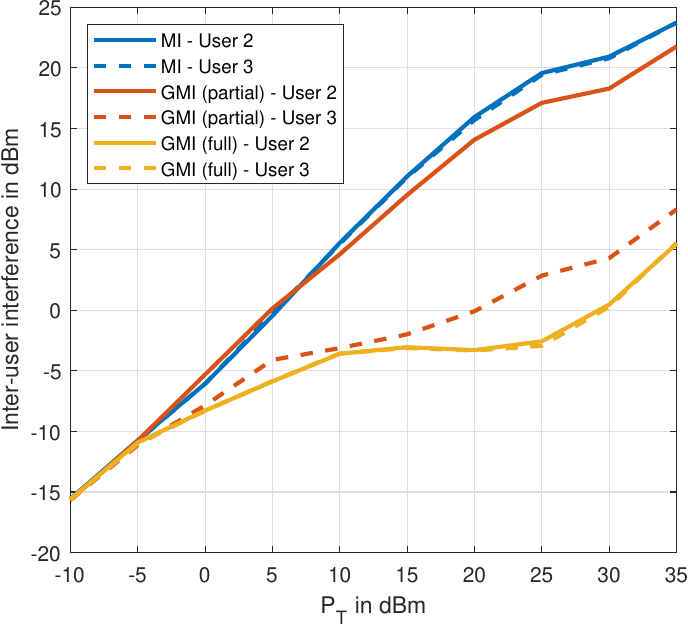}
    \caption{{Ergodic inter-user interference power received at User 1 from the other two users under different objective functions.}}
    \label{Fig: MISO interference power}
\end{figure}

   {To gain further insight into how the precoders are designed differently under various decoding metrics, Figure~\ref{Fig: MISO interference power} depicts the ergodic inter-user interference power received by user-1 from the other two users for different decoding metrics. The results show that the optimized solutions allow interference that is optimally treated to remain strong while keeping suboptimally treated interference weak. This observation is consistent with the results in Section~\ref{section: 2 results}, which indicate that strong interference is not detrimental when optimally treated, but becomes harmful when treated suboptimally.}

\begin{figure}
    \centering
    \includegraphics[width=6.5cm]{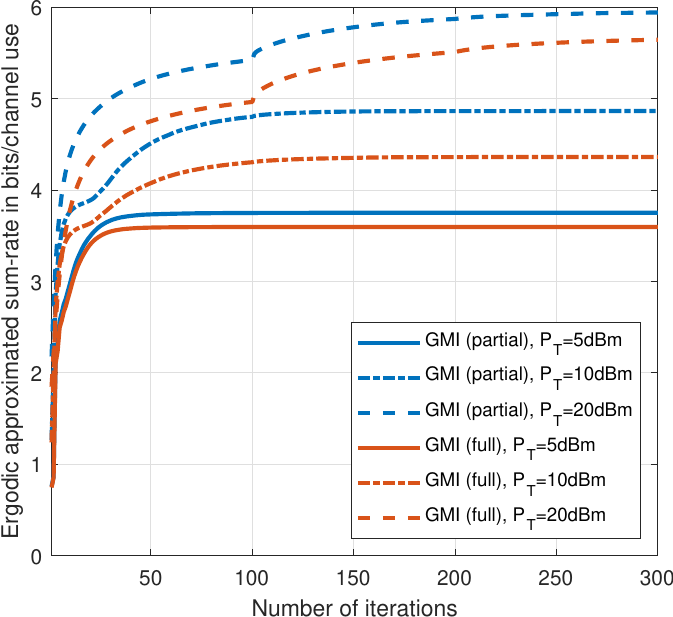}
    \caption{Evolution of the ergodic approximated sum-rate over successive iterations of Algorithm \ref{Algo: GD for SR opt}.}
    \label{Fig: MISO opt convergence}
\end{figure}

   {Figure \ref{Fig: MISO opt convergence} depicts the convergence of the proposed algorithm through the approximated sum-rate obtained at each inner-loop iteration in Algorithm \ref{Algo: GD for SR opt}.}

\subsection{Future Directions}
The above example pioneers a novel system design framework that considers receiver's imperfection in treating finite-alphabet interference, which can be applied to various communication scenarios where interference is critical. Examples of such scenarios include, but are not limited to:
\begin{itemize}
    \item MIMO receivers implemented by linear combiners and single stream decoders, where inter-stream interference may exist during decoding;
    \item Massive random access channels, for example, massive (machine-type) communications in 6G and internet of things, where collision in transmission may happen.
    \item Superposition/non-orthorgonal transmission schemes, where the receivers need to decode under interference brought by the nature of the schemes.
    \item Support for Unmanned Aerial Vehicles (UAV) from cellular networks, where uplink/downklink interference between base stations and UAVs needs to be resolved \cite{3gpp_rel18_WI}.
\end{itemize}
\section{Conclusion}
In this work, we aim at understanding the impact of mismatched/suboptimal decoders under finite-alphabet interference. We first defined the FAGCI model, which takes a finite-alphabet input and adds finite-alphabet interference and Gaussian noise during transmission. We initially considered the optimal decoding metric and two suboptimal decoding metrics, where the suboptimal decoding metrics treat part or all of the interference as Gaussian RV. Achievable rates under different decoding metrics were given by the derived MI and GMI. To improve the achievable rates of FAGCI channel while maintaining low decoding complexity, we further proposed two novel decoding metrics, namely the generalized Gaussian decoding metric and the interference decomposition decoding metric, whose superiority was validated through numerical evaluation on their corresponding GMI. Noting the connection between decoding metrics and demodulators in BICM systems, we also applied the two proposed decoding metrics as demodulators. LLS show that the two novel demodulators also lead to improved throughput. More interestingly, it was also observed that there is a strong correlation between GMI/MI and throughput when the decoding metric corresponds to the demodulator. This indicates that GMI/MI can be used as an efficient tool for evaluating demodulators, and optimizing GMI/MI improves the physical layer throughput under the corresponding demodulator. Lastly, we extended the FAGCI model and the GMI to a multiple-antenna case, and provided an example on MU-MISO precoder optimization, where the objective function is given by MI or GMI, depending on the decoding strategy deployed by the users. The numerical results showed that different decoding strategies led to different throughput and that suboptimal decoders required a dedicated optimization algorithm. Overall, this work provides novel theoretical tools, receiver designs, signal processing framework and insights that can potentially enhance interference management in future wireless networks.
\appendices
\section{Proof of Proposition \ref{matched capacity prop}}\label{Appendix: Proof of matched capacity}
Since uniformly distributed input is assumed, the capacity is given by the mutual information under uniform distribution, i.e., $I_{\text{MI}}(\mathcal{X},\;\mathcal{I},\;\mathcal{J},\;\sigma_z^2) = I(X;\;Y)$. The latter can be expanded using the chain rule of mutual information into
\begin{equation}
\begin{split}
    &I(X;\;Y)\\
    =& I(X,\;I,\;J;\;Y) - I(I,\;J;\;Y \mid X)\\
    =& H(X,\;I,\;J) - H(X,\;I,\;J \mid Y) - H(I,\;J \mid X) + H(I,\;J \mid X,\;Y)\\
    =& \log(|\mathcal{X}|) - H(X,\;I,\;J\mid Y) + H(I,\;J\mid Y-X).
\end{split}
\end{equation}
The last two conditional entropy terms can be computed by definition, resulting in (\ref{matched capacity}).

\section{Proof of Proposition \ref{mismatched capacity prop}}\label{Appendix: Proof of mismatched capacity}
For any mismatched memoryless channel with channel transition probability $f_{Y\mid X}(y\mid x)$ and decoding metric $q(x,y)$, under i.i.d. random coding with input distribution $P_X(x)$, the GMI is given by \cite{Albert}
\begin{equation}\label{general GMI}
\begin{split}
    I_\text{GMI} =& \sup_{s\geq0} \;\; \mathbb{E}_{X,Y} \left[ \log\; \frac{q(X,\;Y)^s}{\mathbb{E}_{\Bar{X}} \left[ q(\Bar{X},\;Y)^s \mid Y)\right ]} \right]\\
    =& \sup_{s\geq0} \;\; s\mathbb{E}_{X,Y} \left[ \log\; q(X,\;Y) \right] - \mathbb{E}_Y \left[\log\; \mathbb{E}_{\Bar{X}} \left[ q(\Bar{X},\;Y)^s \mid Y \right] \right],
\end{split}
\end{equation}
where $(X,\;Y,\;\Bar{X}) \sim P_X(\Bar{x}) P_{Y\mid X}(y\mid x) P_X(x)$.

Under the assumptions given in Section \ref{section: 2.1}, one can obtain
\begin{equation}\label{proof 1.1}
\begin{split}
    &\mathbb{E}_{X,Y}\left[ \log\; q(X,\;Y) \right]\\
    &= \frac{1}{\mid \mathcal{X} \mid \mid \mathcal{I} \mid \mid \mathcal{J} \mid} \sum_{x\in\mathcal{X},\;i\in\mathcal{I},\;j\in\mathcal{J}} \mathbb{E}\left[ \log q(x,\;x+i+j+Z) \right],
\end{split}
\end{equation}
and
\begin{equation}\label{proof 1.2}
\begin{split}
    \mathbb{E}_{\Bar{X}} \left[ q(\Bar{X},\;Y)^s \mid Y \right]
    = \frac{1}{\mid \mathcal{X} \mid} \sum_{\Bar{x}\in\mathcal{X}} q(\Bar{x},\;Y) ^s.
\end{split}
\end{equation}

Based on (\ref{proof 1.2}), (\ref{proof 1.3}) can be obtained. By substituting (\ref{proof 1.1}) and (\ref{proof 1.3}) into (\ref{general GMI}), Proposition \ref{matched capacity prop} is proved.

{To derive the GMI under non-uniformly distributed input signals and/or interference, one only needs to apply the corresponding distribution when evaluating the expectations in (\ref{proof 1.1}), (\ref{proof 1.2}), and (\ref{proof 1.3}).}

\begin{figure*}
\begin{equation}\label{proof 1.3}
\begin{split}
    \mathbb{E}_Y \left[ \log\; \mathbb{E}_{\Bar{X}} \left[ q(\Bar{X},\;Y)^s \mid Y \right] \right]
    = -\log \mid \mathcal{X} \mid + \frac{1}{\mid \mathcal{X} \mid \mid \mathcal{I} \mid \mid \mathcal{J} \mid} \sum_{x\in\mathcal{X},\;i\in\mathcal{I},\;j\in\mathcal{J}} \mathbb{E}_Z \left[ \log \sum_{\Bar{x}\in\mathcal{X}} q(\Bar{x},\;x+i+j+Z) ^s \right]
\end{split}
\end{equation}

\begin{flalign}\label{equ:prop4_1}
    I_{\text{MI}}(\mathcal{X},\;\mathcal{I},\;\mathcal{J},\;\sigma_z^2) = &\log|\mathcal{X}| - \mathbb{E}\left[\log\left( \sum_{\Bar{i}\in\mathcal{I}/I,\;\Bar{x}\in\mathcal{X},\;\Bar{j}\in\mathcal{J}}
    \exp
    \left({\frac{-|X+I+J+Z-\Bar{x}-\Bar{i}-\Bar{j}|^2}{\sigma_z^2}}\right) + \sum_{\Bar{x}\in\mathcal{X},\;\Bar{j}\in\mathcal{J}}\exp
    \left({\frac{-|X+J+Z-\Bar{x}-\Bar{j}|^2}{\sigma_z^2}}\right)\right)\right]&&\nonumber\\
    &+\mathbb{E}\left[\log\left(\sum_{\Bar{i}\in\mathcal{I}/I,\;\Bar{j}\in\mathcal{J}}\exp\left({\frac{-|I+J+Z-\Bar{i}-\Bar{j}|^2}{\sigma_z^2}}\right) + \sum_{\Bar{j}\in\mathcal{J}}\exp\left({\frac{-|J+Z-\Bar{j}|^2}{\sigma_z^2}}\right)\right) \right]&&
\end{flalign}

\begin{flalign}\label{equ:prop4_3}
    \lim_{_{\sigma_i^2\rightarrow 0}}I_{\text{MI}}(\mathcal{X},\;\mathcal{I},\;\mathcal{J},\;\sigma_z^2) = &\log|\mathcal{X}| - \mathbb{E}\left[\log \left(|\mathcal{I}|\sum_{\Bar{x}\in\mathcal{X},\;\Bar{j}\in\mathcal{J}}\exp
    \left({\frac{-|X+J+Z-\Bar{x}-\Bar{j}|^2}{\sigma_z^2}}\right)\right)\right]+\mathbb{E}\left[\log\left(|\mathcal{I}|\sum_{\Bar{j}\in\mathcal{J}}\exp\left({\frac{-|J+Z-\Bar{j}|^2}{\sigma_z^2}}\right)\right) \right]&&\nonumber\\
    =&\log|\mathcal{X}| - \mathbb{E}\left[\log \sum_{\Bar{x}\in\mathcal{X},\;\Bar{j}\in\mathcal{J}}\exp
    \left({\frac{-|X+J+Z-\Bar{x}-\Bar{j}|^2}{\sigma_z^2}}\right)\right]+\mathbb{E}\left[\log\sum_{\Bar{j}\in\mathcal{J}}\exp\left({\frac{-|J+Z-\Bar{j}|^2}{\sigma_z^2}}\right)\right]
    &&\nonumber\\
    =&I_{\text{MI}}(\mathcal{X},\;\{0\},\;\mathcal{J},\;\sigma_z^2)&&
\end{flalign}

\begin{flalign}\label{equ:prop4_5}
    \lim_{\sigma_i^2\rightarrow +\infty}I_{\text{MI}}(\mathcal{X},\;\mathcal{I},\;\mathcal{J},\;\sigma_z^2)  
    =&\log|\mathcal{X}| - \mathbb{E}\left[\log \sum_{\Bar{x}\in\mathcal{X},\;\Bar{j}\in\mathcal{J}}\exp
    \left({\frac{-|X+J+Z-\Bar{x}-\Bar{j}|^2}{\sigma_z^2}}\right)\right]+\mathbb{E}\left[\log\sum_{\Bar{j}\in\mathcal{J}}\exp\left({\frac{-|J+Z-\Bar{j}|^2}{\sigma_z^2}}\right)\right]
    &&\nonumber\\
    =&I_{\text{MI}}(\mathcal{X},\;\{0\},\;\mathcal{J},\;\sigma_z^2)&&
\end{flalign}

{
\begin{flalign}\label{equ:prop 5_1}
    I_\text{GMI}(\mathcal{X},\;\mathcal{I},\;\mathcal{J},\;\sigma_z^2)
    =&\sup_{s\geq 0}\;\; \log|\mathcal{X}|-\mathbb{E} \left[ \log \sum_{\Bar{x}\in\mathcal{X}}\left(\sum_{\Bar{i}\in\mathcal{I}} \exp\left({-\frac{|X + I + J + Z -\Bar{i}-\Bar{x}|^2}{\sigma_j^2+\sigma_z^2}}\right)\right)^s \right]
    + s \mathbb{E}\left[ \log\; \sum_{\Bar{i}\in\mathcal{I}} \exp\left({-\frac{|I + J + Z -\Bar{i}|^2}{\sigma_j^2+\sigma_z^2}}\right)  \right]&&
\end{flalign}}
\hrule
\end{figure*}

\section{Proof of Proposition \ref{prop: MI_GMI_saturation}}\label{Appendix: Proof of MI/GMI saturation}
(\ref{matched capacity}) can be expanded by separating the exponential terms according to whether $I=\Bar{i}$ or not, resulting in (\ref{equ:prop4_1}). If $I\neq\Bar{i}$, $\lim\limits_{\sigma_i^2\rightarrow0} I-\Bar{i} = 0$. Therefore, for terms with $I\neq\Bar{i}$,
\begin{equation}
\begin{split}
    &\lim_{\sigma_i^2\rightarrow0} \exp
    \left({\frac{-|X+I+J+Z-\Bar{x}-\Bar{i}-\Bar{j}|^2}{\sigma_z^2}}\right)\\
    &= \exp
    \left({\frac{-|X+J+Z-\Bar{x}-\Bar{j}|^2}{\sigma_z^2}}\right),
\end{split}
\end{equation}
{and
\begin{equation}
    \lim_{\sigma_i^2\rightarrow0} \exp
    \left({\frac{-|I+J+Z-\Bar{i}-\Bar{j}|^2}{\sigma_z^2}}\right)
    = \exp
    \left({\frac{-|J+Z-\Bar{j}|^2}{\sigma_z^2}}\right),
\end{equation}}
leading to (\ref{equ:prop4_3}). 

From reverse triangle inequality,
\begin{multline}\label{equ:prop4_6}
    \exp
    \left({\frac{-|X+I+J+Z-\Bar{x}-\Bar{i}-\Bar{j}|^2}{\sigma_z^2}}\right)\\
    \leq\exp\left(-\frac{|I-\Bar{i}|^2}{\sigma_z^2}\right)\exp\left({\frac{|X+J+Z-\Bar{x}-\Bar{j}|^2}{\sigma_z^2}}\right).
\end{multline}
If $I\neq\Bar{i}$, 
\begin{equation}
    \lim\limits_{\sigma_i^2\rightarrow+\infty}\exp\left(-\frac{|I-\Bar{i}|^2}{\sigma_z^2}\right)=0.
\end{equation} 
Since the exponential is nonnegative, the squeeze theorem yields
\begin{equation}\label{equ:prop4_7}
\begin{split}
    &\lim_{\sigma_i^2\rightarrow+\infty}\exp
    \left({\frac{-|X+I+J+Z-\Bar{x}-\Bar{i}-\Bar{j}|^2}{\sigma_z^2}}\right) = 0.
\end{split}
\end{equation}
Following a similar approach, it can be proved that
\begin{equation}\label{equ:prop4_8}
\begin{split}
    &\lim_{\sigma_i^2\rightarrow+\infty}\exp\left({\frac{-|I+J+Z-\Bar{i}-\Bar{j}|^2}{\sigma_z^2}}\right) = 0,
\end{split}
\end{equation}
if $I\neq\Bar{i}$. Applying (\ref{equ:prop4_7}) and (\ref{equ:prop4_8}) to (\ref{equ:prop4_1}) results in (\ref{equ:prop4_5}). 

The proof of 
\begin{equation}
\begin{split}
    \lim_{\sigma_i^2\rightarrow0} I_{\text{GMI}}(\mathcal{X,\;I,\;J},\;\sigma_z^2) 
    = &\lim_{\sigma_i^2\rightarrow+\infty} I_{\text{GMI}}(\mathcal{X,\;I,\;J},\;\sigma_z^2)\\
    = &I_{\text{GMI}}(\mathcal{X},\;\{0\},\;\mathcal{J},\;\sigma_z^2)
\end{split}
\end{equation}
with $q(x,y)$ given by (\ref{mismatched decoding metric partial}) follows similar steps, and hence is skipped for clarity.

\section{{Proof of Proposition \ref{prop: GMI_vanish}}}\label{Appendix: Proof of GMI_vanish}
{
With $q(x,\;y)$ given by (\ref{mismatched decoding metric partial}), (\ref{mismatched capacity}) is written as (\ref{equ:prop 5_1}). From reverse triangle inequality,
\begin{multline}\label{equ:prop 5_2}
    \exp\left({-\frac{|X + I + J + Z -\Bar{i}-\Bar{x}|^2}{\sigma_j^2+\sigma_z^2}}\right)\\
    \leq\exp\left({-\frac{|J|^2}{\sigma_j^2+\sigma_z^2}}\right)\exp\left({\frac{|X + I + Z -\Bar{i}-\Bar{x}|^2}{\sigma_j^2+\sigma_z^2}}\right),\\
\end{multline}
and
\begin{equation}
\begin{split}
    &\lim_{\sigma_j^2\rightarrow+\infty} \exp\left({-\frac{|J|^2}{\sigma_j^2+\sigma_z^2}}\right)\exp\left({\frac{|X + I + Z -\Bar{i}-\Bar{x}|^2}{\sigma_j^2+\sigma_z^2}}\right)\\
    &= \exp\left(-\frac{|J|^2}{\sigma_j^2}\right).
\end{split}
\end{equation}
Similarly,
\begin{equation}\label{equ:prop 5_3}
\begin{split}
    &\exp\left({-\frac{|I + J + Z -\Bar{i}|^2}{\sigma_j^2+\sigma_z^2}}\right)\\
    &\leq \exp\left({-\frac{|J|^2}{\sigma_j^2+\sigma_z^2}}\right) \exp\left({\frac{|I + Z -\Bar{i}|^2}{\sigma_j^2+\sigma_z^2}}\right),
\end{split}
\end{equation}
and
\begin{equation}
    \lim_{\sigma_j^2\rightarrow+\infty} \exp\left({-\frac{|J|^2}{\sigma_j^2+\sigma_z^2}}\right) \exp\left({\frac{|I + Z -\Bar{i}|^2}{\sigma_j^2+\sigma_z^2}}\right) = \exp\left(-\frac{|J|^2}{\sigma_j^2}\right).
\end{equation}
Therefore, an upper bound on (\ref{equ:prop 5_1}) follows from (\ref{equ:prop 5_2}) and (\ref{equ:prop 5_3}) and yields
\begin{equation}
\begin{split}
    &\lim_{\sigma_j^2\rightarrow+\infty} I_\text{GMI}(\mathcal{X},\;\mathcal{I},\;\mathcal{J},\;\sigma_z^2) \\
    &\leq \sup_{s\geq 0}\;\; \log|\mathcal{X}|- \mathbb{E}\left[ \log \left(|\mathcal{X}|\left(|\mathcal{I}|\exp\left(-\frac{|J|^2}{\sigma_j^2}\right)\right)^s\right)\right]
    \\
    &\hspace{1cm}+ s\mathbb{E}\left[ \log\left(|\mathcal{I}|\exp\left(-\frac{|J|^2}{\sigma_j^2}\right)\right)\right]\\
    &= \sup_{s\geq 0}\;\; -s\left(\log|\mathcal{I}| + \mathbb{E}\left[ -\frac{|J|^2}{\sigma_j^2}\right]\right) + s\left(\log|\mathcal{I}| + \mathbb{E}\left[ -\frac{|J|^2}{\sigma_j^2}\right]\right)\\
    &= 0.
\end{split}
\end{equation}
Since the rate is nonnegative, we conclude that
\begin{equation}
    \lim_{\sigma_j^2\rightarrow+\infty} I_\text{GMI}(\mathcal{X},\;\mathcal{I},\;\mathcal{J},\;\sigma_z^2) 
    = 0.
\end{equation}
}

\section*{Acknowledgment}
The authors would like to express their sincere gratitude to Prof. Albert Guillén i Fàbregas at the Department of Engineering, University of Cambridge, for the valuable discussions and insights that significantly contributed to the development of this work. Thanks are also due to Dr. David Vargas from BBC Research and
Development for his helpful comments on an earlier draft of this paper.

\bibliographystyle{IEEEtran}
\bibliography{references}

\end{document}